\documentclass{article}
\usepackage{authblk}
\usepackage{cite}
\usepackage{paralist}

\usepackage{psfrag,color}

\usepackage[latin1]{inputenc}
\usepackage{amssymb}
\usepackage{dcolumn,booktabs}
\usepackage{natbib}
\setcitestyle{authoryear,open={(},close={)}}
\usepackage{epsfig}
\usepackage{amsmath}
\usepackage{soul}
\usepackage[usenames,dvipsnames]{xcolor}
\usepackage[left=2cm,right=2cm,top=2cm,bottom=2cm]{geometry}

\providecommand{\keywords}[1]
{
  \small	
  \textbf{\textit{Keywords---}} #1
}

\def\beq{\begin{equation}}
\def\eeq{\end{equation}}
\def\bea{\begin{eqnarray}}
\def\eea{\end{eqnarray}}
\def\cal{\mathcal}

\newcommand{\lp}[1]{#1}

\title{Inflationary Cosmology: From Theory to Observations}
\author[1, 2]{J. Alberto V\'azquez}
\author[2]{Luis, E. Padilla}
\author[2]{Tonatiuh Matos}
\affil[1]{Instituto de Ciencias F\'isicas, Universidad Nacional Aut\'onoma de Mexico, Apdo. Postal 48-3, 62251 Cuernavaca, Morelos, M\'exico. 
}
\affil[2]{Departamento de F\'{\i}sica, Centro de Investigaci\'on y de Estudios Avanzados del IPN, M\'exico.}

\begin{document}

  \maketitle
\vspace{1em}
\begin{abstract}
The main aim of this paper is to provide a qualitative introduction to the cosmological inflation theory
and its relationship with current cosmological observations.
The inflationary model solves many of the fundamental problems that challenge the 
Standard Big Bang cosmology, such as the Flatness, Horizon, and the magnetic Monopole problems. 
Additionally, it provides an explanation for the initial conditions 
 observed throughout the Large-Scale Structure of the Universe, such as galaxies.
In this review, we describe general solutions to the problems in the Big Bang cosmology carry out 
by a single scalar field. Then, with the use of current surveys,
we show the constraints imposed on the inflationary parameters $(n_{\rm s},r)$, which allow us to 
make the connection between theoretical and observational cosmology. 
In this way, with the latest results, it is possible to select, or at least to constrain, the right inflationary model, 
parameterized by a single scalar field potential $V(\phi)$.
\end{abstract}

\begin{abstract}
El objetivo principal de este art\'iculo es ofrecer una introducci\'on cualitativa 
a la teor\'ia de la inflaci\'on c\'osmica y su relaci\'on con observaciones
actuales. El modelo inflacionario resuelve algunos problemas fundamentales que 
desaf\'ian al modelo est\'andar cosmol\'ogico, denominado modelo del Big Bang caliente, 
como el problema de la Planicidad, el Horizonte y la inexistencia de Monopolos magn\'eticos. 
Adicionalmente, provee una explicación al origen de la estructura a gran escala del Universo, como son las galaxias.
En este trabajo se describen soluciones generales \lp{a los problemas de la Cosmolog\'ia del Big Bang} llevadas a cabo por
un campo escalar.  Adem\'as, mediante \lp{observaciones }recientes, se presentan constricciones de los par\'ametros 
inflacionarios  $n_{\rm s}$ y $r$, que permiten realizar la conexi\'on entre la teor\'ia y las observaciones cosmol\'ogicas. De esta manera, con los \'ultimos 
resultados, es posible seleccionar o al menos limitar el modelo inflacionario,  
usualmente parametrizado por un potencial de campo escalar $ V(\phi) $.
\end{abstract}


\section{Introduction}

The Standard Big Bang (SBB) cosmology is currently the most accepted model describing 
the central features of the observed Universe. The Big Bang model, with the addition 
of dark matter and dark energy components, has been successfully proved on 
cosmological levels. For instance, theoretical estimations of the abundance of primordial elements, 
numerical simulations of structure formation of galaxies and galaxy clusters  are in good agreement 
with astronomical observations \citep{Kolbbo, Sping, Planck18}. 
Also, the SBB model predicts the temperature fluctuations observed in the Cosmic 
Microwave Background radiation (CMB) with a high degree of accuracy: 
inhomogeneities of about one part in one hundred thousand \citep{Komat,Planck18}.
These results, amongst many others, are the great success of the SBB cosmology. Nevertheless, 
when we have a closer look at different scales observations seem to present certain 
inconsistencies or unexplained features in contrast  with expected by the theory. 
Some of these unsatisfactory aspects led to the emergence of the inflationary model \citep{Guth, Linde, Linde2, Albrecht}.
\\

In this work, we briefly present some of the relevant shortcomings the standard cosmology is 
dealing with, and a short review is carried out about scalar fields ($\phi$) as promising candidates. 
Moreover, it is shown that an inflationary single canonical-field model can be  completely described 
through its potential energy $V(\phi)$. Also
based on the slow-roll approximation, it is found that the set of parameters that allows 
making the connection with observations is given by the amplitude of density perturbations 
$\delta_H$, the scalar spectral index $n_{\rm s}$, and the tensor-to-scalar ratio $r$.
Finally, the theoretical predictions for different scalar field potentials are shown and 
compared with current observational data on the phase-space parameter $n_{\rm s}-r$, 
therefore pinning down the number of candidates and making predictions about the shape of $V(\phi)$. 
\vspace{1em}

\section{The cosmological model}

\subsection{Main theory}

To avoid long calculations and make this article accessible to young scientists, many 
technical details have been omitted or oversimplified. We encourage the reader to go over the 
vast amount of literature about the inflationary theory
\citep{Lindeb, Kolbbo,  LiddleLyth, Dodelson, Kinney3}.
Before starting  the theoretical description, let us consider some of the assumptions the
SBB model is built \citep{Coles}:
\\ 

1) The physical laws at the present time can be extrapolated further back in time and be
 considered as valid in the early Universe. In this context, gravity is described by
 the theory of General Relativity, up to the Planck era.  
\\

 2) The cosmological principle holds that ``There do not exist preferred places in the Universe";
 that is, the geometrical properties of the Universe over sufficiently large-scales are based 
 on the homogeneity and isotropy, both of them encoded on the Friedmann-Robertson-Walker (FRW) metric

\begin{equation}
 ds^2= -dt^2 + a^2(t)\left[ \frac{dr^2}{1-kr^2} +r^2 \left(d\theta^2 +sin^2\theta\, d\phi^2 \right) \right],
\end{equation}

\noindent
where $(t,r,\theta,\phi)$ describe the time-polar coordinates; the spatial curvature is given by the 
constant $k$, and the cosmic scale-factor $a(t)$ parameterizes the relative expansion of the Universe; 
commonly normalized to today's value $a(t_0)=1$.
Hereafter we use natural units $c=\hbar=1$, where the Planck mass $m_{\rm Pl}$ is related 
to the gravitational constant $G$ through $G\equiv m^{-2}_{Pl}$.
\\
  
 3) On small scales, the anisotropic Universe is described by a linear expansion of the metric around the 
 FRW background:
 
\beq \label{eq:metric}
g_{\mu \nu}(\textbf{x},t)= g_{\mu \nu}^{FRW}(\textbf{x},t)+h_{\mu \nu}(\textbf{x},t).
\eeq

\noindent
To describe the general properties of the Universe, we assume its dynamics are governed by a source treated as a
perfect fluid with pressure $p(t)$ and energy density $\rho(t)$. Both quantities are often related
via an equation-of-state with the form of $p=p(\rho)$. Some of the well studied cases are  

\begin{eqnarray}
p&=& \frac{\rho}{3} \qquad \qquad {\rm Radiation}, \nonumber \\
p&=&0 \qquad \qquad \quad {\rm Dust}, \\
p&=&-\rho  \qquad {\rm Cosmological~ constant}~ \Lambda. \nonumber
\end{eqnarray}

\noindent
The Einstein equations for these kind of constituents, with the FRW metric, are given by the {\bf Friedmann equation}
\beq\label{eq:Friedmann}
H^2  \equiv  \left(\frac{\dot a}{a} \right)^2 = \frac{8\pi}{3 m^2_{\rm Pl}} \rho  - \frac{k}{a^2}, 
\eeq

\noindent
the {\bf acceleration equation}
\beq \label{eq:Acce}
\frac{\ddot{a}}{a}  =   - \frac{4\pi }{3 m^2_{\rm Pl}} (\rho +3p),
\eeq

\noindent
and the energy conservation described by the {\bf fluid equation}

\begin{equation} \label{eq:fluid}
\dot \rho + 3H(\rho + p)=0,
\end{equation}

\noindent
where overdots indicate time derivative, and $H$ defines the \textit{Hubble parameter}. 
Notice that we could get the acceleration equation by time-deriving (\ref{eq:Friedmann}) 
and using (\ref{eq:fluid}); therefore only two of them are independent equations.  
Table \ref{tab:components} displays the solutions for the Friedmann and fluid equations when
different components of the Universe dominate along with the scale factor and the evolution of the 
Hubble parameter in each epoch.
 \\

From Eqn.~(\ref{eq:Friedmann})  can be seen that for a particular 
Hubble parameter, there exists an energy density for which the universe may be spatially flat 
$(k=0)$. This is known as the {\it critical density} $\rho_c$ and is given by
\beq
\rho_c(t)\, =\, \frac{3 m^2_{\rm Pl} \,H^2}{8\pi},
\eeq

\noindent
where $\rho_c$ is a function of time due to the presence of $H$. 
In particular, its current  value is denoted by $\rho_{c,0}=1.87840\, h^2\, \times 10^{-26}$ kg m$^{-3}$, 
or in terms of more  convenient units\lp{,} taking into account large scales in the 
Universe,   $\rho_{c,0}= 2.775 \, h^{-1}\, \times 10^{11} M_{\odot} /(h^{-1} {\rm Mpc})^3 $ 
\citep{Planck18}; with the solar mass denoted by 
$M_{\odot}=1.988\times 10^{33}$g and $h$ parameterizing the present value of the Hubble parameter today

\beq
H_0 = 100 h\, {\rm km\,s}^{-1}{\rm Mpc}^{-1}. 
\eeq

\noindent
The latest value of the Hubble parameter measured by the \textit{Hubble Space Telescope}
is quoted to be \citep{HST}: 
\beq
 H_0= 70.0^{12,0}_{-8,0} \, {\rm km\,s}^{-1} {\rm Mpc}^{-1}. 
\eeq

\begin{table}[t!]
\centering
\begin{tabular}{c|c|c|c}
\toprule
 {\rm component} & \quad $\rho(a)$ & \quad $a(t)$ &\quad $H(t)$ \\	
\hline
 {\rm radiation}	  &\quad $\propto a^{-4}$ & \quad $\propto t^{1/2}$ & \quad 1/(2t)	\\	
{\rm matter}& \quad $\propto a^{-3}$  & \quad $\propto t^{2/3}$  & \quad 2/(3t)	\\	
 cosmological constant & \quad $\propto a^0$ &\quad $\propto \exp(\sqrt{\Lambda \over 3}t)$ & \quad {const} \\
\bottomrule
\end{tabular}
\caption{\footnotesize{E\lowercase{volution of $\rho(a)$, $a(t)$ and} $H(\lowercase{t})$ \lowercase{when 
the Universe is dominated by radiation, matter or a cosmological constant.} }}
\label{tab:components}
\end{table}

At the largest scales a useful quantity to measure is the ratio of the energy density to the critical density
defining the \textit{density parameter} $\Omega_i\equiv \rho_i / \rho_c$. The subscript $i$ labels
different constituents of the Universe, such as  baryonic matter, radiation, dark matter, and dark energy.
The Friedmann equation (\ref{eq:Friedmann}) can be then written such that it
relates the total density parameter and the curvature of the Universe as
\begin{equation} \label{eq:curvature}
 \Omega -1={k \over a^2H^2}.
\end{equation}
Thus the correspondence between the total density content $\Omega$ and the space-time 
curvature for different $k$ values is: 
\begin{itemize}
\item Open Universe : ~$0<\Omega<1: \, k<0: \, \rho<\rho_c$. 
\item Flat Universe       :~ $\Omega=1: \, k=0: \, \rho=\rho_c$. 
\item Closed Universe: $\Omega>1: \, k>0: \, \rho>\rho_c$.
\end{itemize}

\noindent
Current cosmological observations, based on the standard model, find out the present value of 
$\Omega$ is \citep{Planck18} 
\beq \label{eq:Omega}
\Omega_0=1.0007\pm 0.0037,
\eeq
that is, the present Universe is nearly flat.
\\


\subsection{Shortcomings of the model}
This section presents some of the shortcomings the standard old cosmology is facing, to 
then introduce the concept of Inflationary cosmology as a possible explanation to these issues. 
\vskip 10pt

\textbf{Flatness problem}
\vskip 10pt

Notice that $\Omega=1$ is a special case of equation (\ref{eq:curvature}). 
If the Universe was perfectly flat at the earliest epochs,  then it remained so for all time. 
Nevertheless, a flat geometry is an unstable
critical situation; that is, even a tiny deviation from it would cause that $\Omega$ evolved 
quite differently, and very quickly, the Universe would have become more curved. 
This can be seen as a consequence due to $aH$ is a decreasing function of time 
during radiation or matter domination epoch, as it can be observed in Table \ref{tab:components}, then

\bea
&\mid \Omega-1\mid & \, \propto \, t  \hspace{1cm} {\rm during\,\,radiation\,\, domination},\nonumber \\ 
&\mid \Omega-1\mid & \, \propto \, t^{2/3}  \hspace{1cm} {\rm during\,\,dust\,\, domination}. \nonumber
\eea

\noindent
Since the present age of the Universe is estimated to be $t_0 \simeq 13.787$ Gyrs 
\citep{Planck18}, from the above equation, we can 
deduce the required value of $\mid \Omega-1\mid\ =\ \mid \Omega_0-1\mid t/t_0$ at different times to 
obtain the correct spatial-geometry at the present time $\mid \Omega_0-1\mid$ [expression (\ref{eq:Omega})]. 
For instance, let us consider some 
particular epochs in a nearly flat universe:

\begin{itemize}
\item At Decoupling time  $(t \simeq 10^{13}\, {\rm sec})$, we need that $\mid \Omega-1 \mid$ $\le 10^{-3}$.
\item  At Nucleosynthesis time $(t \simeq 1\, {\rm sec})$, we need that $\mid \Omega-1 \mid$ $\le 10^{-16}$.
\item  At the Planck epoch $(t \simeq 10^{-43}\, {\rm sec})$, we need that $\mid \Omega-1 \mid$  $\le 10^{-64}$.
\end{itemize}
Because there is no reason to prefer a Universe with a critical density, hence
$\mid \Omega-1\mid$ should not necessarily be exactly zero. 
Consequently, at early times  $\mid \Omega-1\mid$ has to be fine-tuned extremely close to 
zero to reach its actual observed value.

\vskip 16pt
\textbf{Horizon problem} 
\vskip 10pt
%
The horizon problem is one of the most important problems in the Big Bang model,
as it refers to the communication between different regions of the Universe. 
Bearing in mind the existence of the Big Bang, the age of the Universe is a finite quantity and hence
even light should have only traveled a finite distance by all this time. 

\begin{figure}[t!] 
\centerline{ \epsfxsize=210pt \epsfbox{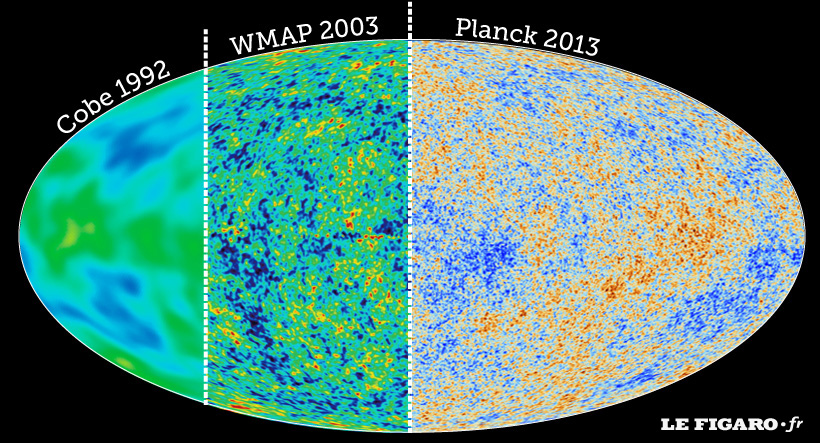} }
\caption{\footnotesize{Temperature fluctuations measured in the CMB radiation using 
 COBE-WMAP-Planck satellites \citep{Gold, Planck18}.}}
\label{fig:wmap5}
\end{figure}

According to the standard cosmology, photons decoupled from the rest of the components at 
temperatures about $T_{dec}\approx 0.3\, eV$ at redshift $z_{dec} \approx 1100$ (\textit{decoupling time}), 
from this time on photons free-streamed and traveled basically uninterrupted until reaching us, giving rise 
to the region known as the \textit{Observable Universe}.
This spherical surface, at which the decoupling process occurred, is called the \textit{surface of the last scattering}.
The primordial photons are responsible for the CMB radiation observed today, then looking at its
fluctuations is analogous of taking a picture of the universe at that time ($t_{dec}\approx 380,000$ years old), 
see Figure \ref{fig:wmap5}.

Figure \ref{fig:wmap5} shows light seen in all directions of the sky, these photons randomly distributed have 
nearly the same temperature $T_0= 2.7255$ K plus small fluctuations (about one part in one hundred thousand) 
\citep{Planck18}.
As we have already pointed out, being at the same temperature is a property of thermal equilibrium. 
Observations are, therefore, easily explained if different regions of the sky had been able to interact and 
moved towards thermal equilibrium. In other words, the isotropy observed in the CMB might imply that 
the radiation was homogeneous and isotropic within regions located on the last scattering surface.
Oddly, the comoving horizon right before  photons decoupled was significantly smaller than the corresponding 
horizon observed today.
This means that photons coming from regions of the sky separated by more than the horizon scale at last 
scattering, typically about $2^\circ$, would not have been able to interact and established thermal equilibrium 
before decoupling. 
A simple calculation displays that at decoupling time, the comoving horizon was 90 $h^{-1}$ Mpc and would 
be stretched up to 2998 $h^{-1}$ Mpc  at present. Then, the volume ratio provides that the microwave 
background should have consisted of  about $\sim10^5$ causally disconnected regions \citep{McCoy}.  
Therefore, the Big Bang model by itself does not explain why
temperatures seen in opposite directions of the sky are so accurately the same; the homogeneity
must had been part of the initial conditions? 
\\

On the other hand, the microwave background is not perfectly isotropic, but instead exhibits
small fluctuations as detected initially by the Cosmic Background Explorer satellite (COBE) \citep{Cobe} and 
 then, with improved measurements, by the Wilkinson Microwave Anisotropy Probe (WMAP)
\citep{wmap5, Larson} and nowadays with the Planck satellite \citep{Planck18}. 
These tiny irregularities are thought to be the `seeds' that grew 
up until becoming the structure nowadays observed in the Universe. 
\\

\vskip 16pt
\textbf{Monopole problem} 
\vskip 10pt

Following the line to find out the simplest theory to describe the Universe, several models in particle physics 
were suggested to unified three out of the  four forces presented in the Standard Model of Particle Physics (SM): 
strong force, described by the group $SU(3)$, weak force, and electromagnetic force, with an associated group $SU(2)\otimes U(1)$. 
These classes of theories are called \textit{Grand Unified Theories (GUT)} \citep{Georgi}.
 An important point to mention in favor of GUT is that they are the only ones that
 predict the equality electron-proton charge magnitude. Also, there are good reasons to 
 believe the origin of \textit{baryon asymmetry} might have been generated on the GUT \citep{Kolb83}.
\\

These kinds of theories assert that in the early stages of the Universe ($t \sim 10^{-43}\, $sec),  at highly extreme 
temperatures ($T_{GUT}\sim 10^{32} \, $K), existed a unified or \textit{symmetric phase} described by a group $G$. 
As the Universe temperature dropped off, it went through different phase transitions until reach 
the symmetries associated with the standard model of particle physics, generating hence
the matter particles such as electrons, protons and neutrons.
When a phase transition happens its symmetry is broken and thus the symmetry group changes by itself,
for instance: 
 \begin{itemize}
 \item GUT transition: $$G \to SU(3)\,\otimes\, SU(2)\, \otimes \, U(1).$$
 \item Electroweak transition: $$SU(3)\,\otimes\, SU(2)\, \otimes \, U(1) \to SU(3)\, \otimes \, U(1).$$ 
\end{itemize}

\noindent
The phase transitions have plenty of implications. One of the most important is the
\textit{topological defects} production which depends on the type of symmetry breaking 
and the spatial dimension \citep{Vilenkin}, some of them are:   

\begin{itemize}
\item Monopoles (zero dimensional).
\item Strings (one dimensional).
\item Domain Walls (two dimensional).
\item Textures (three dimensional).
\end{itemize}

\noindent
Monopoles are therefore expected to emerge as a consequence of unification models. 
Moreover, from particle physics models, there are no theoretical constraints about the mass a monopole should 
carry out. However, from LHC constrictions and grand unification theories, the monopoles would have a 
mass of $10^{13}-10^{18}GeV$ \citep{Mermod}. Hence,  based on their non-relativistic character, 
a crude calculation predicts an extremely high-density number $n_{mono}$ of magnetic 
monopoles ($n_{mono}\sim 10^{76}cm^{-3}$) at the time of grand unified  symmetry breaking \citep{Coles, Ambrosio02}.
%
According to this prediction, the Universe would be dominated by magnetic monopoles.
In contrast with current observations: no one has found anyone yet. 
\\

\section{Cosmological Inflation}
\vskip 6pt

The inflationary model offers the most elegant way so far proposed to solve the problems
of the standard Big Bang and, therefore, to understand the remarkably agreement 
with the standard cosmology. Inflation does not replace the Big Bang model, but rather it is considered 
as an `auxiliary addition', which occurred at the earliest stages of the Universe without disturbing any of its successes.
\\

\textit{Inflation} is defined as the epoch in the early Universe in which the scale factor 
is exponentially expanded in just a fraction of a second:

\bea \label{eq:inflation}
{\rm INFLATION} &\Longleftrightarrow&~~\ddot a>0 \\
&\Longleftrightarrow& \frac{d}{dt}\left( \frac{{1}}{aH}\right)<0. \label{eq:inflation2}
\eea

\begin{figure}[t!] 
\begin{center}
\includegraphics[trim = 1mm  1mm 1mm 1mm, clip, width=7.3cm, height=4.2cm]{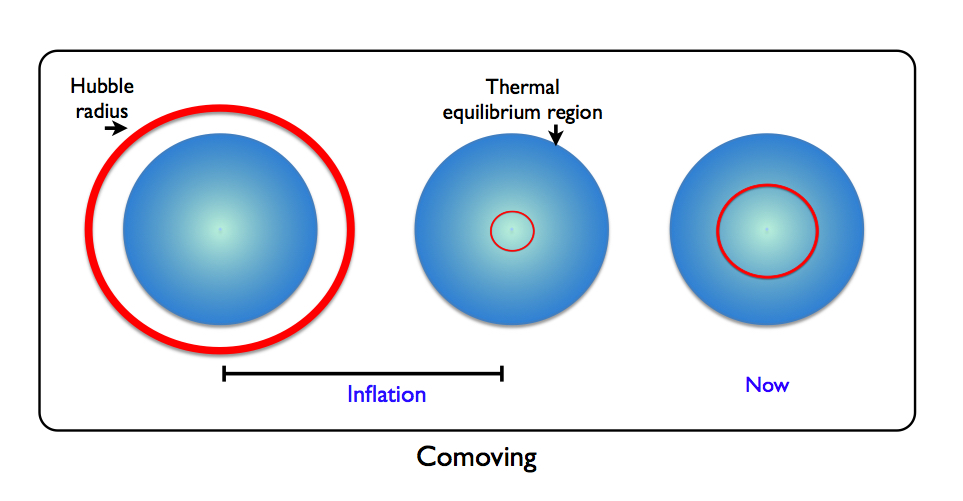}
\includegraphics[trim = 1mm  1mm 1mm 1mm, clip, width=5.6cm, height=4.cm]{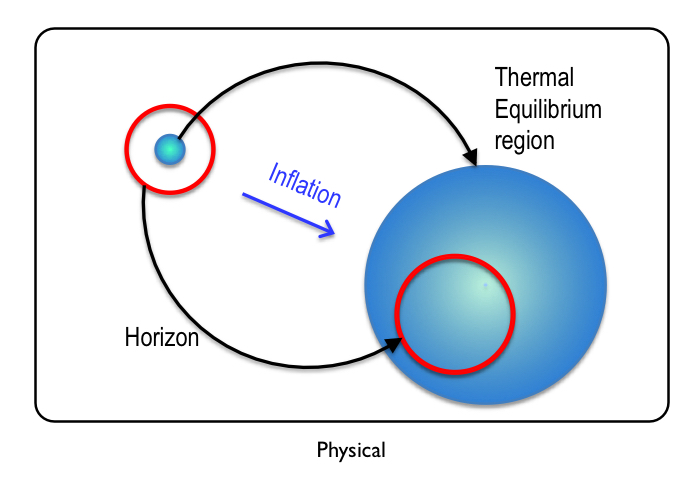}
\caption{\footnotesize{Left: Schematic behavior 
of the comoving Hubble radius during the inflationary period. Right: 
Physical evolution of the observable universe during the inflationary period.}}
\label{fig:Liddle}
\end{center}
\end{figure}

\noindent
The last term corresponds to the comoving Hubble length $1/(aH)$,  interpreted as the observable 
Universe becoming smaller during inflation. This process allowed our observable region to lay down within the Hubble radius at the beginning of inflation. In \citet{Liddle2} words\lp{:} ``is something
similar to zooming in on a small region of the initial universe"; see left panel of Figure \ref{fig:Liddle}.
\\

From the acceleration equation (\ref{eq:Acce}) the condition for inflation, in 
terms of the material required to drive the expansion, is
\beq
\ddot a>0 \Longleftrightarrow (\rho +3p)<0.
\label{gg}
\eeq

\noindent
Because in standard physics it is always postulated $\rho$ as a positive quantity, and hence to satisfy the acceleration 
condition, it is necessary for the overall pressure to have 

\beq 
{\rm INFLATION} ~~ \Longleftrightarrow~~p<-\rho/3.
\eeq

\noindent
Nonetheless, neither a radiation nor a matter component satisfies such condition. 
Let us postpone for a bit the problem of finding a candidate that may satisfy this inflationary condition.

\subsection{Solution for the Big Bang Problems}
If this brief period of accelerated expansion occurred, then the mentioned problems may be solved.
\\

\noindent
\textbf {Flatness problem}
\vskip 6pt
 
A typical solution is a Universe with a cosmological constant $\Lambda$, which can be  interpreted as a perfect 
fluid with equation of state $p=-\rho$. Having this condition, we observe from Table \ref{tab:components} that  
the universe is exponentially expanded:
\beq
a(t)\propto \exp\left(\sqrt{\frac{\Lambda}{3}}t\right),
\eeq
and the Hubble parameter $H$ is constant, then the condition  (\ref{eq:inflation2}) is naturally fulfilled. 
This epoch is called \textit{de Sitter stage}. However, postulating a cosmological constant as a candidate to 
drive inflation might create more problems than solutions by itself, i.e., reheating process \citep{Carrol01}.
\\

Let us look at what happens when a general solution is considered. If somehow there was an accelerated expansion, 
$1/(aH)$ tends to be smaller on time, and hence, by the expression (\ref{eq:curvature}), $\Omega$ is driven 
towards the unity rather than away from it. 
Then, we may ask ourselves how much should $1/(aH)$ decrease. If the inflationary period started at time $t=t_i$ 
and ended up approximately at the beginning of the radiation dominated era ($t=t_f$), then 

\begin{figure}[t!] 
\centerline{ \epsfxsize=200pt \epsfbox{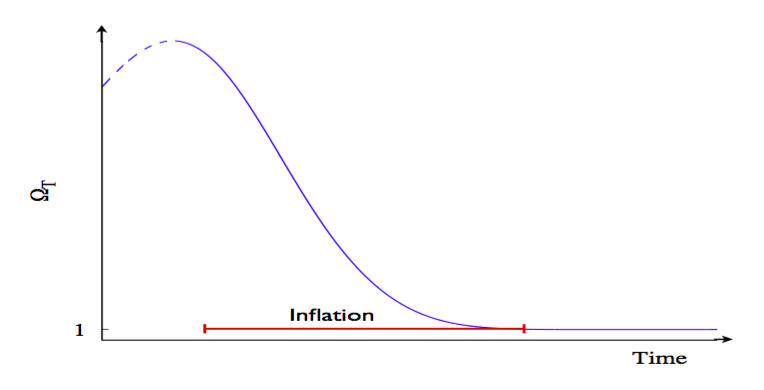} }
\caption{\footnotesize{Evolution of the density parameter $\Omega$ during the inflationary period. $\Omega$ is
driven towards unity, rather than away from it.}}
\label{fig:curvature}
\end{figure}

$$
\mid \Omega -1\mid_{t=t_f}\sim10^{-60},
$$
and
\beq
\frac{\mid \Omega -1\mid_{t=t_f}}{\mid \Omega -1\mid_{t=t_i}}= \left( a_i \over a_f \right)^2\equiv e^{-2N}.
\eeq
\\

So, the required condition to reproduce the value of $\Omega_0$  measured today is that inflation lasted for at 
least $N\equiv \ln a \gtrsim 60 $, then $\Omega$ must be extraordinarily close to one that we still observe such quantity today.
In this sense, inflation magnifies the curvature radius of the universe, so 
locally the universe seems to be flat with great precision, Figure \ref{fig:curvature}.

\vskip 16pt
\noindent
\textbf{Horizon problem}
\vskip 6pt
%
As we have already seen, during inflation, the universe expands drastically, and there is a  reduction in the comoving 
Hubble length. This process allowed a tiny region located inside the Hubble radius to evolve and constitute our present 
observable Universe.  Fluctuations were hence stretched outside of the horizon during inflation and re-entered the horizon 
in the late Universe, see Figure \ref{fig:Liddle}. Scales outside the horizon at CMB-decoupling were, in fact, inside the 
horizon before inflation. The region of space corresponding to the observable universe, therefore, was in thermal equilibrium 
before inflation, and the uniformity of the CMB is essentially explained.
\\

\vskip 16pt
\noindent
\textbf{Monopole problem}
\vskip 6pt
The monopole problem was initially the main motivation to develop the inflationary
cosmology \citep{Guth}.
During the inflationary epoch, the Universe led to a dramatic expansion over which the density of the unwanted particles 
were diluted away. Generating enough expansion, the dilution made sure the particles stayed completely out of 
the observable Universe making pretty difficult to localize even a single magnetic monopole.     


\section{Single-field inflation}
\vskip 6pt

Throughout the literature, there exists a broad diversity of models that have been suggested to carry out the inflationary process 
\citep{LiddleLyth, Olive, Lyth}. In this section, we present the scalar fields as good candidates  to drive inflation and explain 
how to relate theoretical predictions to observable quantities.  Here, we limit ourselves to models based on general 
gravity, i.e., derived from the Einstein-Hilbert action, and single-field models described by a homogeneous 
real slow-rolling scalar field $\phi$. 
Nevertheless, in section \ref{sec:multi} we provide a very brief introduction to inflation with several scalar fields, as a possibility
to generate the inflationary process.
\\

Inflation relies on the existence of an early epoch in the Universe dominated by a very different form of energy; remember the 
requirement of the unusual negative pressure. Such a condition can be satisfied by a single scalar field (spin-0 particles). 
The scalar field, which drives the Universe to an inflationary epoch, is often termed as the \textit{inflaton field}. 

Let us consider a real scalar field minimally coupled to gravity, with an arbitrary
potential $V(\phi)$ and Lagrangian density $\mathcal{L}$ specified by the action

\begin{equation}
S=\int d^4x\, \sqrt{-g}\,\mathcal{L}=\int\, d^4x\, \sqrt{-g}\,
\left[\frac{1}{2}
\partial_{\mu}\phi
\partial^{\mu}\phi -V(\phi)\right].
\end{equation}
The energy-momentum tensor corresponding to this field is given by
\beq
T_{\mu\nu}=\partial_{\mu}\phi \partial_{\nu}\phi
-g_{\mu\nu}\, \mathcal{L}.
\eeq
In the same way as the perfect fluid treatment, the
energy density $\rho_\phi$ and pressure density $p_\phi$ in the FRW metric are found to be 
\begin{eqnarray}
T_{00}=\rho_{\phi}=\frac{1}{2}\dot{\phi}^2 + V(\phi)+ \frac{1}{2}\nabla \phi^2,  \\
T_{ii}=p_{\phi}=\frac{1}{2}\dot{\phi}^2 - V(\phi)- \frac{1}{6}\nabla \phi^2.
\end{eqnarray}

\noindent
Considering a homogeneous field ($\nabla \phi=0$), its corresponding equation of state is
 
\begin{equation}
w = \frac{p_\phi}{\rho_\phi}=\frac{\frac{1}{2}\dot \phi^2-V(\phi)}{\frac{1}{2}\dot \phi^2+V(\phi)}.
\end{equation}

\noindent
We can now split up the inflaton field as
\beq \label{eq:split}
\phi({\bf x},t)=\phi_{0}(t)+\delta\phi({\bf x},t),
\eeq
where $\phi_{0}$ is considered a classical field, that is, 
the mean value of the inflaton on the homogeneous and isotropic state, 
whereas $\delta\phi({\bf x},t)$ describes the quantum fluctuations around $\phi_{0}$.

\noindent
The evolution equation for the background field $\phi_0$  is given by
\begin{equation}
\ddot{\phi_0}+ 3H\dot{\phi_0}= -V_{,\phi_0}.
\label{eq:motion1}
\end{equation}

\noindent
Moreover, the Friedmann equation (\ref{eq:Friedmann}) with negligible curvature becomes

\beq \label{eq:motion2}
H^2 = \frac{8\pi}{3m^2_{\rm Pl}} \left[{1 \over 2} \dot\phi_0^2 +V(\phi_0)\right],
\eeq
where we have used commas as derivatives with respect to the scalar field $\phi_0$. 
\\

 From the structure of the effective energy density and pressure, the acceleration
  equation (\ref{eq:Acce}) becomes, 
 
 \beq
 {\ddot a \over a} = -{8\pi \over 3m_{\rm Pl}^2}\left(\dot \phi_0^2-V(\phi_0) \right).
 \eeq
 
 \noindent
 Therefore, the inflationary condition to be satisfied is $\dot \phi_0^2 < V(\phi_0)$, which 
 is easily fulfilled with a suitably flat potential. Now, we shall omit the subscript
 `0' by convenience.

\subsection{Slow-roll approximation}
\vskip 6pt

As we have noted, a period of accelerated expansion can be created by 
the cosmological constant $(\Lambda)$ and hence solve the aforementioned problems.
After a brief period of time, inflation must end up, and its energy converted into conventional
matter/radiation; this process is called \textit{reheating}. In a universe dominated by a 
cosmological constant, the reheating process is seen as $\Lambda$ decaying into 
conventional particles; however, claiming that $\Lambda$ is able to decay is still a 
naive way to face the problem.   
On the other hand, scalar fields have the property to behave like a 
\textit{dynamical cosmological constant}. Based on this approach, it is useful to
suggest a scalar field model starting with a nearly flat potential, i.e., initially 
satisfies the \textit{first slow-roll} condition $\dot \phi^2 \ll V(\phi)$. 
This condition may not necessarily be fulfilled for a long time, but
to avoid this problem, a second \textit{slow-roll} condition is defined as 
$|\ddot{\phi}|\ll |V,_{\phi}|$ or equivalently $|\ddot{\phi}|\ll 3H|\dot{\phi}|$. In this case, the scalar field is slowly rolling 
down its potential, and by obvious reasons, such approximation is called \textit{slow-roll} 
\citep{Liddle92, Liddle94}.
The equations of motion (\ref{eq:motion1}) 
 and (\ref{eq:motion2}), for slow-roll inflation, then become
 
\bea \label{eq:slow}
3H\dot{\phi} ~~ &\simeq& ~~ -V_{,\phi}, \\
H^2 ~~ & \simeq& ~~ \frac{8\pi}{3m^2_{\rm Pl}} V(\phi). \label{eq:slow2}
\eea

\noindent
It is easily verifiable that the slow-roll approximation requires the slope 
and curvature of the potential to be small: $V_{,\phi}, V_{,\phi\phi} \ll V$.
\\

\begin{figure}[t!] 
\centerline{ \epsfxsize=170pt \epsfbox{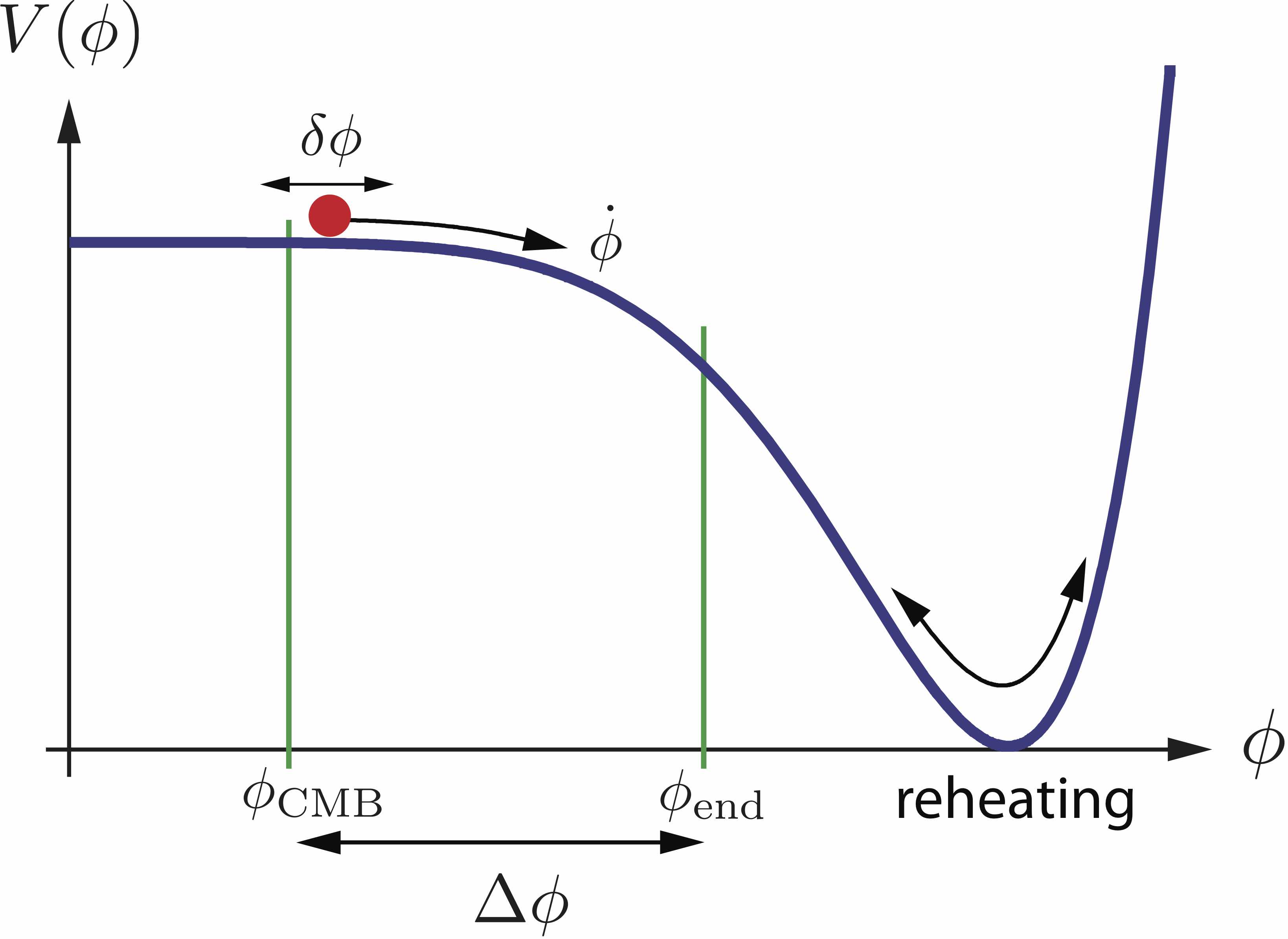} }
\caption{\footnotesize{Schematic inflationary process \citep{Baumann2}.}}
\label{fig:Field}
\end{figure}

The inflationary process happens when the kinetic part of the inflaton field is subdominant over the potential field $V(\phi)$. 
When both quantities become comparable, the inflationary period ends up giving 
rise finally to the reheating process, see Fig.~\ref{fig:Field}. 

\vspace{0.5cm}
It is now useful to introduce the \textit{potential slow-roll parameters} 
$\epsilon_{\rm v}$ and $\eta_{\rm v}$ in the following way \citep{Liddle92, Riotto17}

\bea
\epsilon_{\rm v}(\phi) &\equiv&{m^2_{\rm Pl} \over 16 \pi } \left({V_{,\phi} \over V}\right)^2, 
\label{eq:epsi} \\
\eta_{\rm v}\left(\phi\right) &\equiv& \frac{m^2_{\rm Pl}}{8\pi} {V_{,\phi\phi } \over V} \label{eq:eta}.
\eea
Equations (\ref{eq:slow}) and (\ref{eq:slow2}) are in agreement with the slow-roll approximation
when the following conditions hold

\begin{equation*}
\epsilon_{\rm v}(\phi) \ll 1,  \,\,\,\,\,\,  \mid \eta_{\rm v}(\phi)\mid \ll 1.
\end{equation*}

\noindent
These conditions are sufficient but not necessary because the validity of the slow-roll
approximations was a requirement in its derivation.
The physical meaning of $\epsilon_{\rm v}(\phi)$ can be explicitly seen by expressing equation (\ref{eq:inflation})
 in terms of $\phi$, then, the inflationary condition is equivalent to
 
 \begin{equation}
 {\ddot a \over a} ~~ > ~~ 0 ~~ \Longrightarrow ~~ \epsilon_{\rm v}(\phi) < ~~  1.
\end{equation}

\noindent
Hence, inflation concludes when the value $\epsilon_{\rm v} (\phi_{end})= 1$ is reached.
\\

Within these approximations, it is straightforward to find out the scale factor between the beginning and the 
end of inflation. Because the size of the expansion is an enormous quantity, it is useful to compute it in terms of the 
 {\it e}-fold number $N$, defined by 
 
\begin{equation} \label{eq:N}
N \equiv \ln {a(t_{end}) \over a(t)}=
\int_{t}^{t_e}{H\,dt} \simeq 
{8\pi \over m^2_{\rm Pl}} \int_{\phi_e}^{\phi} {V \over V_{,\phi}} d\phi .
\end{equation}

\noindent
To give an estimate of the number of \textit{e}-folds, let assume the evolution of the Universe 
can be split up into different epochs and concentrate on a particular scale $k$ (at this point, we only consider a generic scale; 
however, in the next section we will explain that such scales can be associated to the size of perturbations in a Fourier space), 
which was inside the horizon at the beginning of inflation and then at certain time left the horizon. 
If we consider particularly the moment when the size of such scale was equal to the horizon, i.e., $k=aH$, then we can 
assume the following cosmological history:

\begin{itemize}
\item Inflationary era: horizon crossing ($k=aH$) $\to$ end of inflation $a_{end}$.
\item Radiation era: reheating $a_{reh}$ $\to$ matter-radiation equality $a_{eq}$.
\item Matter era: $a_{eq}$ $\to$ present $a_{0}$.
\end{itemize}

\noindent
Assuming the transition between one era to another is instantaneous, then $N(k)= \ln ({a_k / a_0})$
can be easily computed with:
$$ 
{k\over a_0 H_0}\,=\,{a_k H_k \over a_0 H_0}\,=\,{a_k\over a_{end}}{a_{end}\over a_{reh}}
{a_{reh}\over a_{eq}}{a_{eq} \over a_0}{H_k\over H_0},
$$
where $a_k$ $(H_k)$ refers to the scale factor (Hubble parameter) measured at the moment when 
$k$ equals the horizon. Then, one has \citep{LiddleLyth}

$$
N(k)=62-\ln{k \over a_0 H_0}-\ln{10^{16} GeV \over V_k^{1/4}}+\ln{V_k^{1/4} \over V_{end}}-
{1 \over 3}\ln{V_{end}^{1/4} \over \rho_{reh}^{1/4}}.
$$
The last three terms are small quantities related to energy scales during the inflationary process and usually can be ignored.
The precise value for the second quantity depends on the model as well as the Planck normalization; however, it does 
not present any significant change to the total amount of \textit{e}-folds. 
Thus, the value of total \textit{e}-foldings is ranged from 50-70 \citep{Lyth}. 
Nevertheless, this value could change if a modification of the full history of the Universe is considered. 
For instance, thermal inflation can alter $N$ up to a minimum value of $N=25$ \citep{Lyth1,Lyth2}.
\\

As we noted, the parameters describing inflation can be presented as a function of the scalar field potential. That is, an inflationary 
model with a single scalar field is specified by selecting an inflationary potential $V(\phi)$. 
At this point, it is necessary to mention that these potentials are not chosen arbitrarily,  but in fact, there is a whole line of research 
motivated by fundamental physics. For this paper, we will not delve into this subject; however, 
it will be understood that this potential is motivated by some fundamental theory. To exemplify our initial point, 
let us consider the following example.

The potential that describes a massive \lp{and free} scalar field is given by:

\beq \label{eq:mass}
V(\phi)= \frac{1}{2}m^2 \phi^2.
\eeq

\noindent
Considering the slow-roll approximation, equations (\ref{eq:motion1}) and (\ref{eq:motion2}) become:
\bea
3H\dot \phi &=& -m^2 \phi, \\
H^2 &=& \frac{4\pi m^2 \phi^2}{3 m_{\rm Pl}^2}.\nonumber
\eea
\noindent
Thus, the dynamics of this type of model is described by

\bea
\phi(t)&=&\phi_i - \frac{m m_{\rm Pl}}{\sqrt{12 \pi}}t,\\
a(t)&=& a_i \exp\left[\sqrt{\frac{4\pi}{3}}\frac{m}{m_{\rm Pl}}\left( \phi_i t - \frac{m m_{\rm Pl}}{\sqrt{48 \pi}}t^2 \right) \right], \nonumber
\eea

\noindent
where $\phi_i$ and $a_i$ represent the initial conditions at a given initial time $t=t_i$.
The slow-roll parameters for this particular potential are computed from equations (\ref{eq:epsi}) and (\ref{eq:eta})

\beq
\epsilon_{\rm v}=\eta_{\rm v}= \frac{m_{\rm Pl}^2}{4\pi} \frac{1}{\phi^2}, 
\eeq

\noindent
that is, an inflationary epoch takes place while the condition $|\phi|> {m_{\rm Pl}}/\sqrt{4\pi}$ 
is satisfied, and the total amount lapsed during this accelerated period is encoded on the $e$-folds number

\beq\label{N_chaotic}
N_{tot}= \frac{2\pi}{m_{\rm Pl}^2}\left[\phi^2_i - \phi^2_e \right].
\eeq 

The steps shown before might, in principle, apply to any inflationary single-field model. 
That is, the general information we need to characterize \lp{the} 
cosmological inflation is specified by \lp{the scalar field potential responsible for generating this mechanism}.

\subsection{Cosmological Perturbations}

Inflationary models have the merit that they do not only explain the  homogeneity of the Universe on large-scales
but also provide a theory  for explaining the observed level of {\em anisotropy}. During the inflationary period, quantum 
fluctuations of the field were driven to scales much larger than the Hubble horizon. Then, in this process, the fluctuations
were frozen and turned into metric perturbations \citep{Mukhanov}. 
Metric perturbations created during inflation can be described by two terms. The {\it scalar, or curvature,} perturbations 
are coupled with matter in the Universe and form the initial ``seeds'' of structure observed in galaxies today. 
Although the {\it tensor perturbations} do not couple to matter, they are associated to the generation of primordial 
gravitational waves. As we shall see, scalar and tensor perturbations are seen as important components to 
 the CMB anisotropy \citep{Hu}. 
\\

In a similar matter we introduced the density parameter for large scales, on small scales, we consider the 
\textit{density contrast} defined by $\delta \equiv \delta \rho / \rho$. At this point, it is convenient to work in a Fourier 
description and then quantities are replaced by its corresponding analog in Fourier space, for example, 
$\delta(\text{\textbf{x}},t)\rightarrow \delta_{\textbf{k}}(\textbf{k},t)$, where \textbf{k} refers to a given scale, 
and similarly for several quantities, and $k = |\textbf{k}|$.
We now on assume \textit{adiabatic initial conditions}, which require that matter and radiation perturbations 
are initially in perfect thermal equilibrium, and therefore the
density contrast for different species in the Universe satisfy 

 \beq
 {1 \over 3}\delta_{{\bf k} b}={1 \over 3}\delta_{{\bf k} c}={1 \over 4}\delta_{{\bf k} \gamma}
 \left(={1 \over 4}\delta_{{\bf k} }\right),
 \eeq 

\noindent
where subindex $\text{\textbf{k}}b,\text{\textbf{k}}c,\text{\textbf{k}}\gamma$ refer to the density contrast in Fourier space 
for baryons, dark matter, and radiation, respectively, and $\delta_k$ is the total density contrast. 
We encourage 
the reader to look at \citep{LiddleLyth} or \citep{peebles} for a more accurate description of the above important relation. 
The most general density perturbation is described by a linear combination of adiabatic
perturbations as well as \textit{isocurvature perturbations}, \lp{where} the latter one plays an important role 
when more than one scalar field is considered (see next section and \citet{LiddleLyth}).
\\

On the other hand, the \textit{primordial curvature perturbation} $\mathcal{R}_k(t)$ has the property 
to be constant within a few Hubble times after the horizon exit, i.e. when $k=aH$.  
This value is called the \textit{primordial value} and is related to the scalar field perturbation $\delta \phi_k$ by

\beq\label{eq:Pk1}
\mathcal{R}_k=-\left[ {H \over \dot \phi}\,\, \delta \phi_k \right]_{k=aH}.
\eeq

 \noindent
 As already mentioned, if inflation provides an exponential expansion, then the horizon remains practically constant 
 while all other scales grow up. In this way, we can focus on the evolution of the quantum perturbations of the inflaton 
 into a small region compared to the horizon. In this region, it is possible to assume the space as locally flat and ignore the 
 metric perturbations. Thus, working in Fourier space the classical equation of motion for the perturbation part of 
 $\phi({\bf x},t)$ in (\ref{eq:split}) is
\beq
(\delta \phi_k)\ddot \,\,+3 H (\delta \phi_k)\dot \,\,+\left({k \over a}\right)^2 \delta \phi_k=0,
\eeq
where we have assumed linear perturbations and neglect higher orders. This means that perturbations generated by
vacuum fluctuations have uncorrelated Fourier modes, the signature of \textit{Gaussian perturbations}. 
\\

The above equation can be rewritten as a harmonic oscillator equation with variable frequency. If we now move to 
the quantum world and make the corresponding associations of operators to classical variables, the quantum dynamics 
will be determined  by  \citep{LiddleLyth2}

\begin{equation}\label{qscalarfield}
\hat{\psi}_k\left(\eta\right)=\frac{\psi_k\left(\eta\right)\hat{a}\left(k\right)+\psi^*_k\left(\eta\right)\hat{a}^\dagger\left(-k\right)}{\left(2\pi\right)^3} \ \ \ \text{with} \ \ \ \psi_k\left(\eta\right)=-\frac{e^{-ik\eta}}{\sqrt{2k}}\frac{k\eta-i}{k\eta},
\end{equation}
where $\hat{a}$ and $\hat{a}^\dagger$ are the particle creation and annihilation 
operators, $\eta$ is the conformal time defined by $\partial_\eta\equiv a\partial_t$, where during inflation 
$\eta \sim -1/aH$ and $\psi\equiv a\delta\phi$.
\\

The inflationary process dilutes all possible particles existing before this period. Taking this into account,  the ground 
state of the system is given by the vacuum. 
We notice that well after horizon exit, $\eta \rightarrow 0$, $\psi_k\left(\eta\right)$ approaches the value
\begin{equation} \label{psi_k}
\psi_k\left(\eta\right)=-\frac{i}{\sqrt{2k}}\frac{1}{k\eta},
\end{equation}
so that equation (\ref{qscalarfield}) is rewritten as
\begin{equation}
\hat{\psi}_k\left(\eta \right)=\psi_k\left(\eta\right)\frac{\hat{a}\left(k\right)-\hat{a}^\dagger\left(-k\right)}{\left(2\pi\right)^3}.
\end{equation}
The temporal dependence of $\hat{\psi}_k$ is now trivial and implies that once $\psi_k\left(\eta\right)$ is measured after horizon 
exit, it will continue having  a definite value. This quantum fluctuation becomes classical once the horizon is crossed  and can 
be taken as the initial inhomogeneity that will later give rise to the  structure formation. 
However, these initial conditions will be slightly modified due to the amount of inflation remaining, once the $k$-scale has left the horizon.
\\

Defining the spectrum of perturbations as 
\bea \label{eq:two_point}
\langle\psi_k\psi^*_{k'}\rangle &=& \frac{2\pi^2}{k^3} \mathcal{P}_\psi (k)\delta_D (\vec{k}-\vec{k}'),  \nonumber \\
 					&=&\frac{2\pi^2}{k^3}a^2 \mathcal{P}_\phi (k)\delta_D (\vec{k}-\vec{k}'),
 \eea
where the Dirac's delta distribution $\delta_D$ guarantees that modes relative to different wave-numbers are 
uncorrelated to preserve homogeneity. \lp{In the above expression, the quantity $P_{\phi}$ ($P_{\psi}$) is the spectrum 
generated by the perturbed part of the field $\phi$ ($\psi=a\delta \phi$)}. The left-hand side of the equation (\ref{eq:two_point})
(along with the expression (\ref{psi_k})) evaluated at a few Hubble times after the horizon exit, 
$\eta \sim 1/aH_k$, 
yields to the spectrum
\beq\label{eq:Pk2}
\cal P_{\phi}(k)=\left( {H \over 2\pi }\right)^2_{k=aH}.
\eeq
From (\ref{eq:Pk1}) and (\ref{eq:Pk2})
 the  primordial curvature power spectrum $\cal P_{\cal R}(k)$, computed in terms of the scalar
 field spectrum $\cal P_{\phi}(k)$, is given by 
\begin{eqnarray}\label{eq:pspectrum}
\cal P_{\cal R}(k) &=& \left[\left({H \over \dot \phi} \right)^2 \cal P_{\phi}(k) \right]_{k=aH}  \nonumber \\
			   &= &\left[ \left({H \over \dot \phi}\right) \left({H \over 2 \pi}\right)\right]^2_{k =a H} .
\end{eqnarray}

On the other hand, the creation of primordial gravitational waves corresponds to the tensor 
part of the metric perturbation $h_{\mu \nu}$ in (\ref{eq:metric}). In Fourier space, 
tensor perturbations $h_{ij}$ can be expressed as the superposition of two polarization modes
\beq
h_{ij}= h_{+}\mathit{e}^{+}_{ij}+h_{\times}\mathit{e}^{\times}_{ij},
\eeq
where $+$, $\times$ represent the longitudinal and transverse modes. From Einstein equations, it is found that each 
amplitude $h_{+}$ and $h_{\times}$ behaves as a free scalar field in the sense that
\beq
\psi_{+,\times}\equiv {m_{\rm Pl} \over \sqrt 8}\,\,h_{+,\, \times}.
\eeq 

\noindent
Therefore, taking the results of the scalar perturbations, each $h_{+,\, \times}$ has a spectrum $\cal P_T$ given by

\beq
\cal P_{T}(k)={8 \over m_{\rm Pl}^2} \left({H \over 2\pi} \right)_{k=aH}^2.
\eeq
The canonical normalization of the field $\psi_{+,\times}$ was chosen such that
the \textit{tensor-to-scalar ratio} of the spectra is  

\begin{equation}\label{tensortoscalar}
r \equiv {\cal P_{T} \over \cal P_{\cal R}} =16 \epsilon_{\rm v}.
\end{equation}

During the horizon exit, $k=aH$, $H$ and $\dot \phi$ have tiny variations during a few Hubble times. In this case, the 
scalar and tensor  spectra are nearly scale-invariant and therefore well approximated to a power law 
\bea \label{eq:spectra}
\cal P_\cal R(k)  =\cal P_\cal R(k_0) \left( \frac{k}{k_0} \right)^{n_{\rm s}-1},\qquad 
\cal P_T(k)  = \cal P_T(k_0) \left( \frac{k}{k_0} \right)^{n_T} \, .
\eea

\noindent
where $k_0 = 0.002$Mpc$^{-1}$ and the spectral indices are defined as 

\beq
n_{\rm s}-1\equiv {d\ln \cal P_\cal R(k) \over d\ln k},\qquad
n_T \equiv {d\ln \cal P_T(k) \over d\ln k}. 
\eeq

\noindent
A scale-invariant spectrum, called Harrison-Zel'dovich (HZ), has constant variance on all length scales, and it is 
characterized by $n_{\rm s}= 1$; small deviations from scale-invariance are also considered as a typical 
signature of the inflationary models.  Then the spectral indices $n_{\rm s}$ and $n_{T}$ can be expressed 
 in terms of the slow-roll parameters $\epsilon_{\rm v}$ and $\eta_{\rm v}$, to lowest order, as:

\bea \label{indices}
n_{\rm s} - 1  &\simeq&  - 6~ \epsilon_{\rm v}(\phi) + 2~ \eta_{\rm v}(\phi), \nonumber \\
n_T  &\simeq&  -2~ \epsilon_{\rm v}(\phi). 
\eea

\noindent
These parameters are {\em not} completely independent,  but the tensor spectral index is proportional to 
the tensor-to-scalar ratio $r = -8 n_{T}  $.  This expression is the \textit{first consistency relation} for slow-roll inflation.  
Hence, any inflationary model, to the lowest order in slow-roll, can be described in terms of three independent parameters:
 the amplitude of density perturbations  $\delta \sim \cal P_\cal R(k_0)^{1/2}$ ($\approx 5 \times 10^{-5}$ initially measured 
 by COBE satellite), the scalar spectral index $n_{\rm s}$, and the tensor-to-scalar ratio $r$. If we require a more accurate 
 description, we have to consider higher-order effects, and then include parameters for describing the running of 
 scalar ($n_{\rm s_{run}} \equiv d n_{\rm s} / d\ln{k}$), tensor ($n_{T_{run}} \equiv  d n_T / d\ln{k}$) index, and higher 
 order corrections. 
\\

An important point to emphasize is that  $\delta$, $r$, and $n_{\rm s}$ are  parameters that nowadays are tested from 
several observations. This allows comparing theoretical predictions with observational data,  for instance, those provided 
by the Cosmic Microwave Background radiation. In other words, future measurements of these parameters may
probe or at least constrain the inflationary models, and therefore the shape of the inflaton potential $V(\phi)$.
\\

Let us get back to the massive-free scalar field example in equation (\ref{eq:mass}).
Inflation ends up when the condition $\epsilon_{\rm v}=1$ is achieved, so $\phi_{end}\simeq m_{\rm Pl}/\sqrt{2 \pi}$.
 As we pointed out before, we are interested in models with an $e$-fold number of about $N_{tot}=60$, that is from \eqref{N_chaotic}
 \beq
 \phi_i=\phi_{60}\simeq \sqrt{\frac{30}{\pi}}m_{\rm Pl}.
 \eeq
 
 \noindent
 Finally, the spectral index and the tensor-to-scalar ratio for this potential are
 
 \beq
 n_{\rm s}-1= -\frac{1}{30}, \qquad r= \frac{2}{15}.
 \eeq

\noindent
If the massive scalar field potential is the right inflationary model, current observations should favor
the values $n_{\rm s}\approx 0.97$ and $r\approx 0.1$.
\\

To determine the shape of the primordial power spectrum [Eqn. (\ref{eq:pspectrum})] from cosmological observations, it is 
usual to assume a parameterized form for it. Even though the simplest assumption for the spectra has a form of a 
power-law given by Eqn. (\ref{eq:spectra}), there have been several studies regarding the shape of the
primordial spectrum. Some of them based on physical models, some using observational data to constrain an a priori parameterization, 
and others attempting a direct reconstruction from data \citep{Vazquez1, Hlozek,  Vazquez2, Guo, Vazquez3}

\section{Multi-field inflation}\label{sec:multi}

Assuming that a single scalar field is responsible for inflation may be only an approximation 
since the presence of multiple fields could drive this process as well. In this section, we show how the cosmological equations 
are modified when two scalar fields are responsible for driving the inflationary process  \citep{Byrnes}. 
 The generalization of several fields can be easily obtained and described by  \citep{Gong}. 

\subsection{Background equation of motion}

We consider a two-field inflationary model with canonical kinetic terms and dynamics described by an arbitrary interaction 
potential $V(\phi, \psi)$.  As usual, we assume the classical fields are homogeneous and evolve in an FRW background. 
Thus, the background equation of motion for each scalar field and the Hubble parameter are

\begin{subequations}\label{eq:56}
\begin{equation}\label{KGEq}
\ddot{\phi}_i+3H\dot{\phi_i}+\frac{dV_i}{d|\phi_i|^2}\phi_i=0, \quad (i=\phi,\psi),
\end{equation}

\begin{equation}
H^2=\frac{8\pi}{3m_{\rm Pl}^2}\left[V+\frac{1}{2}\left(\dot\phi^2+\dot\psi^2\right)\right],
\end{equation}
\end{subequations}
where $V_{,i}\equiv\partial V/\partial \phi_i$. During inflation, we adopt the slow-roll approximation for each field. 
This occurs always that the condition $\epsilon_i,|\eta_{ij}|\ll 1$ is fulfilled; $\epsilon_i$ and $\eta_{ij}$ are now 
a new set of slow-roll parameters defined by
\begin{equation}\label{slowroll2f}
\epsilon_i = \frac{m_{\rm Pl}^2}{16 \pi }\left(\frac{V_{,i}}{V}\right)^2, \ \ \ \ \ \ \ \ 
\eta_{ij}=\frac{m_{\rm Pl}^2}{8\pi}\left(\frac{V_{,ij}}{V}\right).
\end{equation}
The set of equations \eqref{eq:56} are rewritten in the slow-roll approximation as
\begin{equation}
\dot{\phi}_i\simeq -\frac{V_{,i}}{3H}\left(1+\frac{1}{3}\delta_i^H\right), \ \ \ \ \ \ \ \ \
H^2\simeq \frac{8\pi }{3m_{\rm Pl}^2}V\left(1+\frac{1}{3}\epsilon^H\right)
\end{equation}
with $\delta_i^H$ and $\epsilon^H$ the new slow-roll parameters:
\begin{equation}
\delta_i^H = -\frac{\ddot\phi_i}{H\dot\phi_i}, \ \ \ \ \ \ \ \ \epsilon^H = \epsilon_{\phi\phi}+\epsilon_{\psi\psi}.
\end{equation} 

\subsection{Cosmological perturbations: the adiabatic and isocurvature perturbations}

The equation of motion for each perturbed field is described by

\begin{equation}
\ddot{\delta\phi}_i+3H\dot{\delta\phi}_i+\sum_j\left[V_{,ij}-\frac{8\pi}{a^3m_{\rm Pl}^2}\frac{d}{dt}\left(\frac{a^3}{H}\dot{\phi}_i \dot\phi_j\right)\right]\delta\phi_j=0.
\end{equation}

\noindent
On the largest scales ($k\ll aH$) it is better to work on a rotating basis of the fields defined by the relation:
  \begin{subequations}
  \begin{equation}
  \binom{\delta \sigma}{\delta s}=S^{\dagger}\binom{\delta \phi}{\delta\psi},
  \end{equation}
  where
  \begin{equation}\label{angle}
  S=\begin{pmatrix}\cos\theta & -\sin\theta\\ \sin\theta & \cos\theta\end{pmatrix}, \ \ \ \tan\theta =\frac{\dot \psi}{\dot \phi}\simeq\pm \sqrt{\frac{\epsilon_\psi}{\epsilon_\phi}}.
  \end{equation} 
  \end{subequations}
The field $\sigma$ is parallel to the trajectory in field space, and it is usually called 
the \textit{adiabatic field}, whereas the field  $s$ is perpendicular, named the \textit{entropy field}. 
If the background trajectory is curved, then $\delta\sigma$ and $\delta s$ are correlated at Hubble exit,
and therefore, at such moment, the \textit{power spectra} and \textit{cross-correlation} are described by the expressions:

\begin{subequations}
\begin{equation}
\left.\cal P_{\sigma}(k)\right|_{k=aH}\simeq\left(\frac{H}{2\pi}\right)^2_{k=aH}(1+(-2+6C)\epsilon-2C\eta_{\sigma\sigma}),
\end{equation}
\begin{equation}
\left.C_{\sigma s}(k)\right|_{k=aH}\simeq-2C\eta_{\sigma s}\left(\frac{H}{2\pi}\right)^2_{k=aH},
\end{equation}
\begin{equation}\label{5c}
\left.\cal P_{s}(k)\right|_{k=aH}\simeq\left(\frac{H}{2\pi}\right)^2_{k=aH}(1+(-2+2C)\epsilon-2C\eta_{ss}),
\end{equation}
\end{subequations}
where $C\simeq 0.7296$, $\epsilon\equiv \epsilon_{\sigma\sigma}+\epsilon_{ss}$ and $\eta_{ij}$ ($i,j=\sigma,s$) are 
slow-roll parameters defined in a similar way than Eq. \eqref{slowroll2f}, but now in terms of the new fields $\sigma$ and $s$.

\subsubsection{Final power spectrum and spectral index}

The \textit{curvature} and \textit{isocurvature perturbations} are usually defined as
\begin{equation}\label{RS}
\cal R\equiv\frac{H}{\dot\sigma}\delta \sigma, \ \ \ S=\frac{H}{\dot \sigma}\delta s.
\end{equation}
In the slow-roll limit, on large scales, the evolution of curvature and isocurvature 
perturbations can be written using the formalism of transfer matrix:
\begin{equation}
\binom{\cal R }{S}=\begin{pmatrix}1 & T_{\cal RS}\\ 0& T_{SS}\end{pmatrix}\binom{\cal R}{S}_{k=aH},
\end{equation}
where
\begin{equation}
T_{SS}(t_k,t)=\exp\left(\int^t_{t_k}\beta Hdt'\right), \ \ \ \ \ \ 
T_{\cal RS}(t_k,t)=\exp\left(\int^t_{t_k}\alpha T_{SS}Hdt'\right),
\end{equation}
being $t_k$ the time at horizon crossing. At linear order in slow-roll parameters
\begin{equation}
\alpha\simeq -2\eta_{\sigma s}, \ \ \ \ \beta\simeq-2\epsilon+\eta_{\sigma\sigma}-\eta_{ss},
\end{equation}
where again $\eta_{ij}$ is defined similarly than Eqs. \eqref{slowroll2f} but in terms of the new fields $\sigma$ and $s$.

\noindent
On the other hand, the primordial curvature perturbation during the radiation-dominated 
era (some time after inflation finished) is given, on large scales, by
\begin{equation}
\cal R=\Psi+\frac{H\delta\rho}{\rho},
\end{equation}
where $\Psi$ is the gravitational potential. The conventional definition of the isocurvature perturbation 
for an $i$-specie is given relative to the radiation density by
\begin{equation}
S_i=H\left(\frac{\delta\rho_{i}}{\rho_{i}}-\frac{\delta\rho_\gamma}{\rho_\gamma}\right).
\end{equation}
Then, at the beginning of the radiation-domination era, we get the final power spectra 
\begin{subequations}\label{spectrums}
\begin{equation}\label{PRf}
\cal P_\cal R\simeq P|_{k=aH}(1+\cot^2\Delta),
\end{equation}
\begin{equation}\label{isosecond}
\cal P_S=T^2_{SS}P|_{k=aH},
\end{equation}
\begin{equation}
C_{\cal RS}=T_{RS}T_{SS}P_R|_{k=aH},
\end{equation}
\end{subequations}
where at linear order in slow-roll parameters $P|_{k=aH}$ is 
\begin{equation}
P|_{k=aH}=\frac{1}{2\epsilon}\left(\frac{2H}{m_{\rm Pl}}\right)^2_{k=aH},
\end{equation}
with $\Delta$ the observable correlation angle defined at the lowest order by
\begin{equation}
\cos\Delta =\frac{T_{RS}}{\sqrt{1+T_{RS}^2}}.
\end{equation}

\noindent
The final spectral index for each contribution, defined as $n_x-1=d\ln P_x/d\ln k$, at linear order in slow-roll parameters, are
\begin{subequations}\label{tilts}
\begin{eqnarray}
n_{\rm s}-1&\simeq & -(6-4\cos^2\Delta)\epsilon+2\sin^2\Delta\eta_{\sigma\sigma},\nonumber \\ 
&&+4\sin\Delta\cos\Delta\eta_{\sigma s}+2\cos^2\Delta\eta_{ss},\\
n_{\rm S}-1&\simeq & -2\epsilon+2\eta_{ss}.\\
n_C-1&\simeq &-2\epsilon+2\tan\Delta\eta_{\sigma s}+2\eta_{ss},
\end{eqnarray}
\end{subequations}
Notice that we have kept the subindex $\rm s$ to be consistent with the scalar spectral index defined in 
the single inflationary scenario.
It is also common to parameterize the primordial adiabatic and entropy perturbations on super-horizon 
scales as power laws
\begin{subequations}\label{PswAs}
\begin{equation}\label{PrAs}
\cal P_\cal R=A_r^2\left(\frac{k}{k_0}\right)^{n_{ad1}-1}+A_s^2\left(\frac{k}{k_0}\right)^{n_{ad2}-1},
\end{equation}
\begin{equation}\label{PrCrs}
C_{\cal RS}=A_sB\left(\frac{k}{k_0}\right)^{n_{cor}-1},
\end{equation}
\begin{equation}\label{PsAs}
\cal P_S=B^2\left(\frac{k}{k_0}\right)^{n_{iso}-1},
\end{equation}
\end{subequations}
where at linear order $n_{ad1}=-6\epsilon+2\eta_{\sigma\sigma}$, $n_{ad2}=2n_C-n_{\rm S}$, $n_{cor}=n_C$, $n_{iso}=n_{\rm S}$. We have that $A_r^2$, $A_s^2$ and $B$ can be written in terms of the correlation angle as
\begin{subequations}
\label{RelAs}
\begin{equation}
A_r^2=[\cal P_\cal R\sin^2\Delta]_{k_0}, \ \ \ \ A_s^2=[\cal P_\cal R\cos^2\Delta]_{k_0},
\end{equation}
\begin{equation}
B^2=[T_{SS}^2 \cal P_\cal R|_*]_{k_0},
\end{equation}
\end{subequations}
$A_r^2$ and $A_s^2$ are the contributions of the adiabatic and entropy fields to the amplitude of the primordial adiabatic spectrum. 

\subsubsection{Gravitational waves}

Given the fact that scalar and tensor perturbations are decoupled at linear order, 
 gravitational waves at horizon crossing are the same as in the single-field case. Also,
 their amplitude should remain frozen on large scales after Hubble exit. Therefore 
 the tensor power spectrum and the spectral index are finally
\begin{equation}
\cal P_T=\left.\cal P_{T}\right|_{k=aH}\simeq 8 \left(\frac{H}{2\pi m_{\rm Pl}}\right)^2_{k=aH}(1+2(-1+C)\epsilon),
\end{equation}
\begin{equation}\label{tiltsnt}
n_T\simeq -2\epsilon\left[1+\left(\frac{4}{3}+4C\right)\epsilon+\left(\frac{2}{3}+2C\right)\eta_{\sigma\sigma}\right],
\end{equation}

\noindent
The tensor-to-scalar ratio at Hubble exit is the same as in the single field case. However, at super-horizon scales, 
the curvature perturbations continue evolving as \eqref{PRf}. In this way the value of $r$ some time after the end of inflation is
\begin{equation}\label{Tensortoscalar}
r\simeq 16\epsilon \sin^2\Delta\left[1-\left(\frac{4}{3}+4C\right)\epsilon +\left(\frac{2}{3}+2C\right)\eta_{\sigma\sigma}\right].
\end{equation}
We can observe from \eqref{tensortoscalar} that the single scalar field case works as an upper constraint on $r$.

\section{Inflationary models}

We have seen that a single-field inflationary model could be described by the specification of the potential form $V(\phi)$. 
In this case, the comparison of  model predictions to CMB observations reduces to the following basic steps:
\begin{enumerate}
 \item Given a scalar field potential $V(\phi)$, compute the slow-roll parameters $\epsilon_{\rm v}(\phi)$ and
$\eta_{\rm v}(\phi)$. 
\item Find out $\phi_{end}$ given by $\epsilon_{\rm v}(\phi_{end})=1$. 
\item From equation (\ref{eq:N}), compute the field at about 60 $e$-folds $\phi_{60}$.
\item  Compute $n_{\rm s}$ and $r$ as function of $\phi_{60}$ \lp{to test the model with CMB data}. 
\end{enumerate}

Different types of models are classified by the relationship amongst their slow-roll parameters $\epsilon_{\rm v}$ 
and $\eta_{\rm v}$, which are reflected in different relations between $n_{\rm s}$ and $r$. Hence, an appropriate 
parameter space to show the diversity of models  is well described by the $n_{\rm s}$---$r$ plane.

\subsection{Models}

\begin{figure}[t!] 
\begin{center}
  \includegraphics[trim = -50mm 1mm 10mm 1mm, clip, width=10cm, height=5cm]{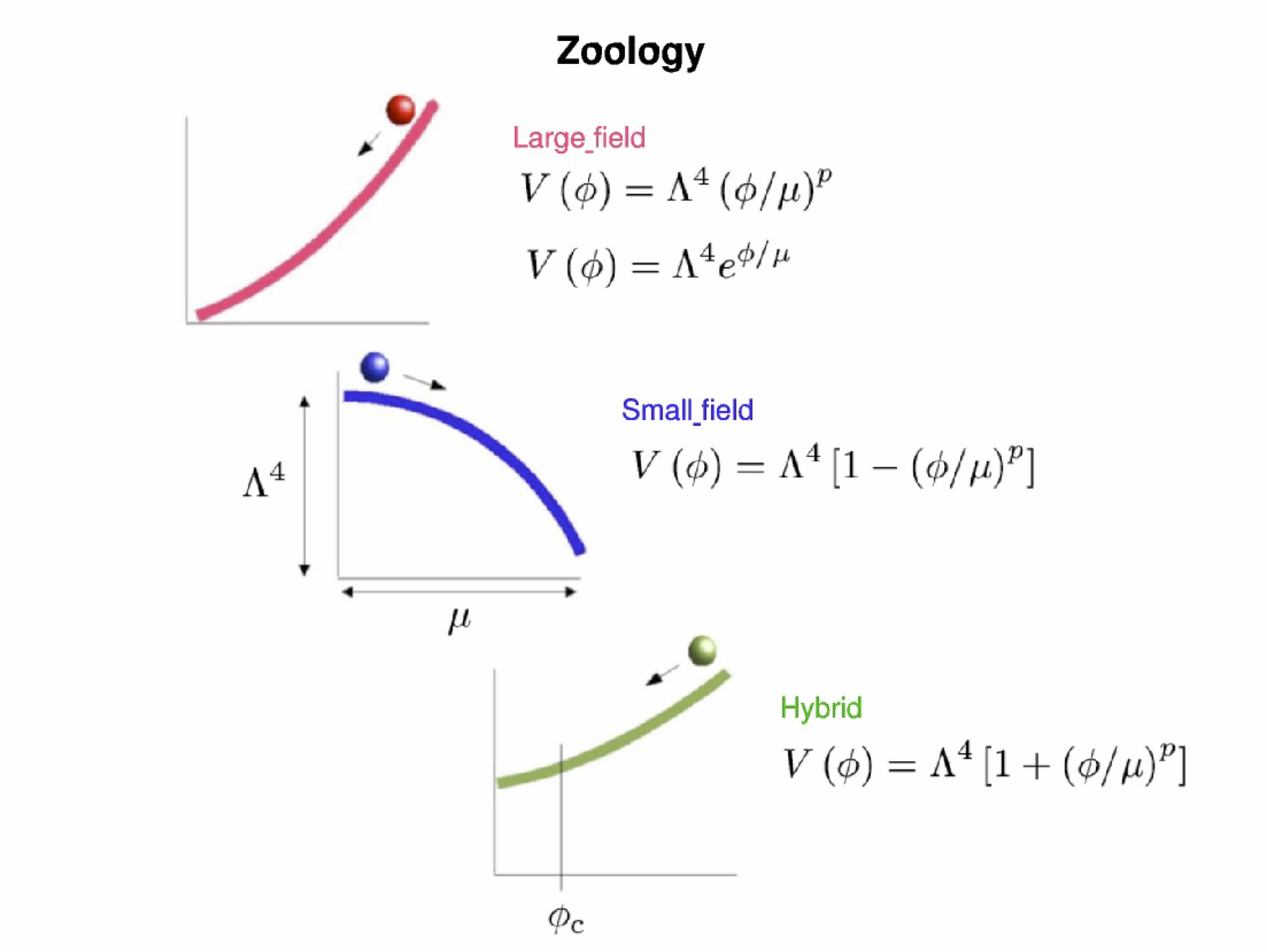}
\caption{\footnotesize{Potential classification. From top to bottom:
\textit{large field, small field, and hybrid potential} \citep{Kinney3}.}}
\label{fig:models}
\end{center}
\end{figure}

Even if we restrict the analysis to a single-field, the number of inflationary models available is enormous 
\citep{LiddleLyth, Lyth, Linde05, Kinney2}. Then, it is convenient to classify different kinds of potentials following \citet{Kinney2}. 
The classification is based on the behavior of the potential during inflation. The three basic types are shown in Figure \ref{fig:models}.
{\em Large field}: the field is initially displaced from a stable minimum and evolves 
towards it. {\em Small field}: the field evolves away from an unstable maximum. 
{\em Hybrid}: the field evolves towards a minimum with vacuum energy different from zero. 

\noindent
A general single field potential can be written in terms of a \textit{height} $\Lambda$ and a 
\textit{width} $\mu$, such as
\begin{equation}
V\left(\phi\right) = \Lambda^4 f\left({\phi \over \mu}\right).
\end{equation}
Different models have different forms for the function $f$.

\subsection{Large-field models: $-\epsilon_{\rm v} < \eta_{\rm v} \leq \epsilon_{\rm v}$}

\begin{figure}
 \begin{center}
  \includegraphics[trim = 20mm 120mm 10mm 40mm, clip, width=7cm, height=4cm]{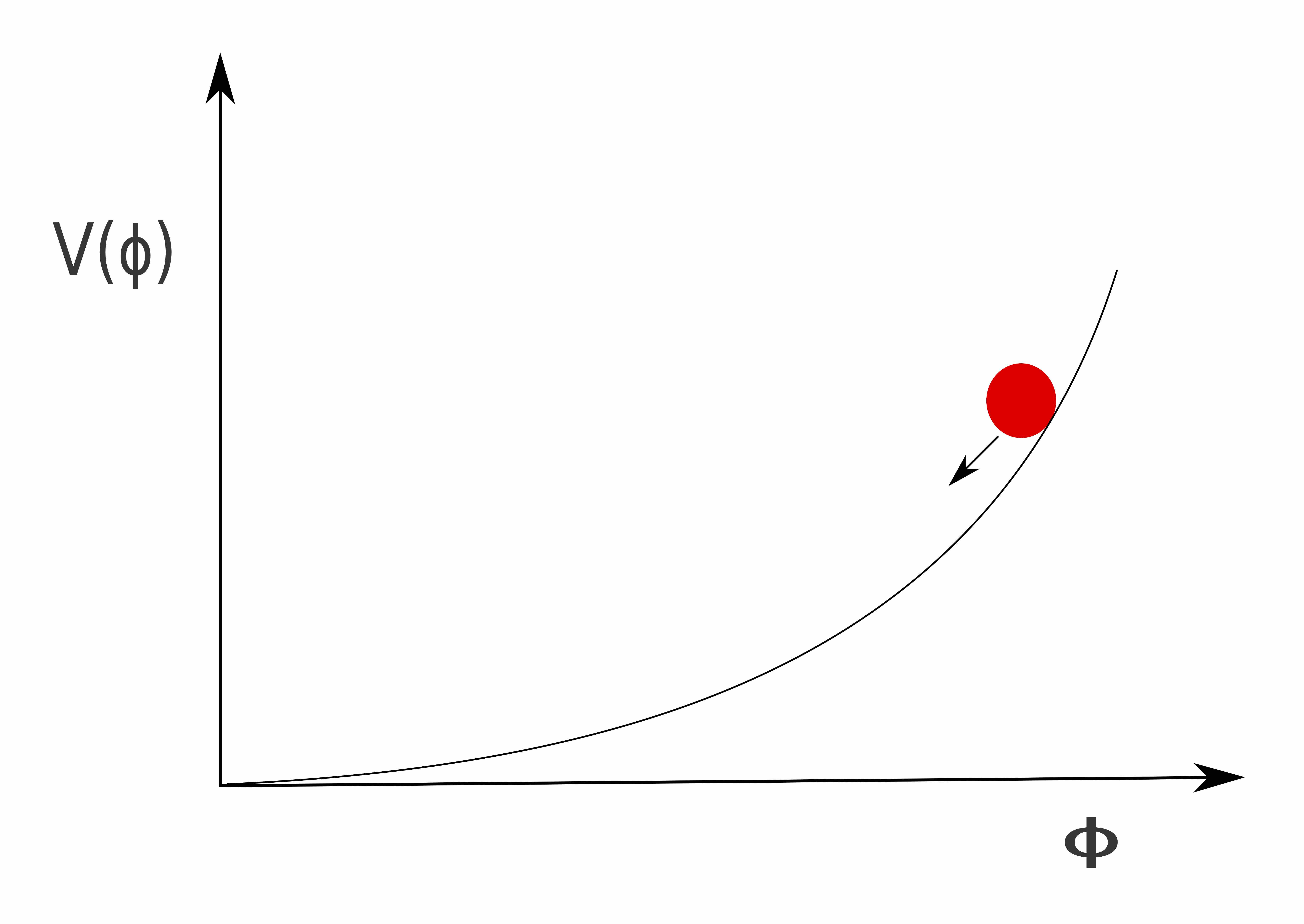}
	\caption{\footnotesize{Chaotic inflationary potential.}}
	\label{fig:new1}
\end{center}	
\end{figure}

\textbf{Large field} models perhaps posses the simplest type of monomial potentials. These kinds of potentials represent 
the \textit{chaotic} inflationary scenarios \citep{Linde2}. The distinctive of these models is that the  shape of the effective 
potential is not very important in detail. That is, a region of the Universe where the scalar field is usually situated at 
$ \phi \sim m_{\rm Pl}$  from the minimum of its potential will automatically lead to inflation (Figure \ref{fig:new1}). 
Such models are described by $V_{,\phi\phi} > 0$ and $-\epsilon_{\rm v} < \eta_{\rm v}
\leq \epsilon_{\rm v}$. 

\noindent
 A general set of large-field polynomial potentials can be written as
\begin{equation}
V\left(\phi\right) = \Lambda^4 \left({\phi \over \mu}\right)^p,
\end{equation}
where it is enough to choose the exponent $p>1$ in order to specify a particular model.
This model gives 
\begin{eqnarray}\label{nsr}
n_{\rm s}-1&=&-\frac{2+p}{2N}\, ,\nonumber\\
r&=&\frac{4 p}{N}.
\end{eqnarray}

\noindent
In this case, gravitational waves can be sufficiently big to eventually be observed $(r\gtrsim 0.1)$.
From the quadratic potential of equation (\ref{eq:mass}), we obtain

\begin{equation}
\epsilon_{\rm v} \simeq 0.008, \quad \eta_{\rm v} \simeq 0.008, \quad n_{\rm s} \simeq 0.97, \quad r \simeq 0.128.
\end{equation}

\noindent
In the high power limit, the $V \propto \phi^p$ predictions are the same as the exponential 
potential \citep{Steinhardt}. Hence, a variant of this class of models is 

\begin{equation}
V\left(\phi\right) = \Lambda^4 \exp\left(\phi / \mu\right).
\end{equation}
 This type of potential is a rare case presented in inflation because its dynamics has an exact solution given by a power-law expansion.
 For this case the spectral index $n_{\rm s}$ is closely related to the tensor-to-scalar ratio $r$, as 
\begin{eqnarray}
n_{\rm s}-1&=&-\frac{m^2_{\rm Pl}}{8\pi \mu^2}, \nonumber \\
r &=& 8 \left(1 - n_{\rm s}\right),
\end{eqnarray}
as we observe, the slow-roll parameters are explicitly independent of the \textit{e}-fold number $N$. 

\subsection{Small-field models: $\eta_{\rm v} < -\epsilon_{\rm v}$}

\begin{figure}
 \begin{center}
  \includegraphics[trim = 20mm 120mm 10mm 40mm, clip, width=7cm, height=4cm]{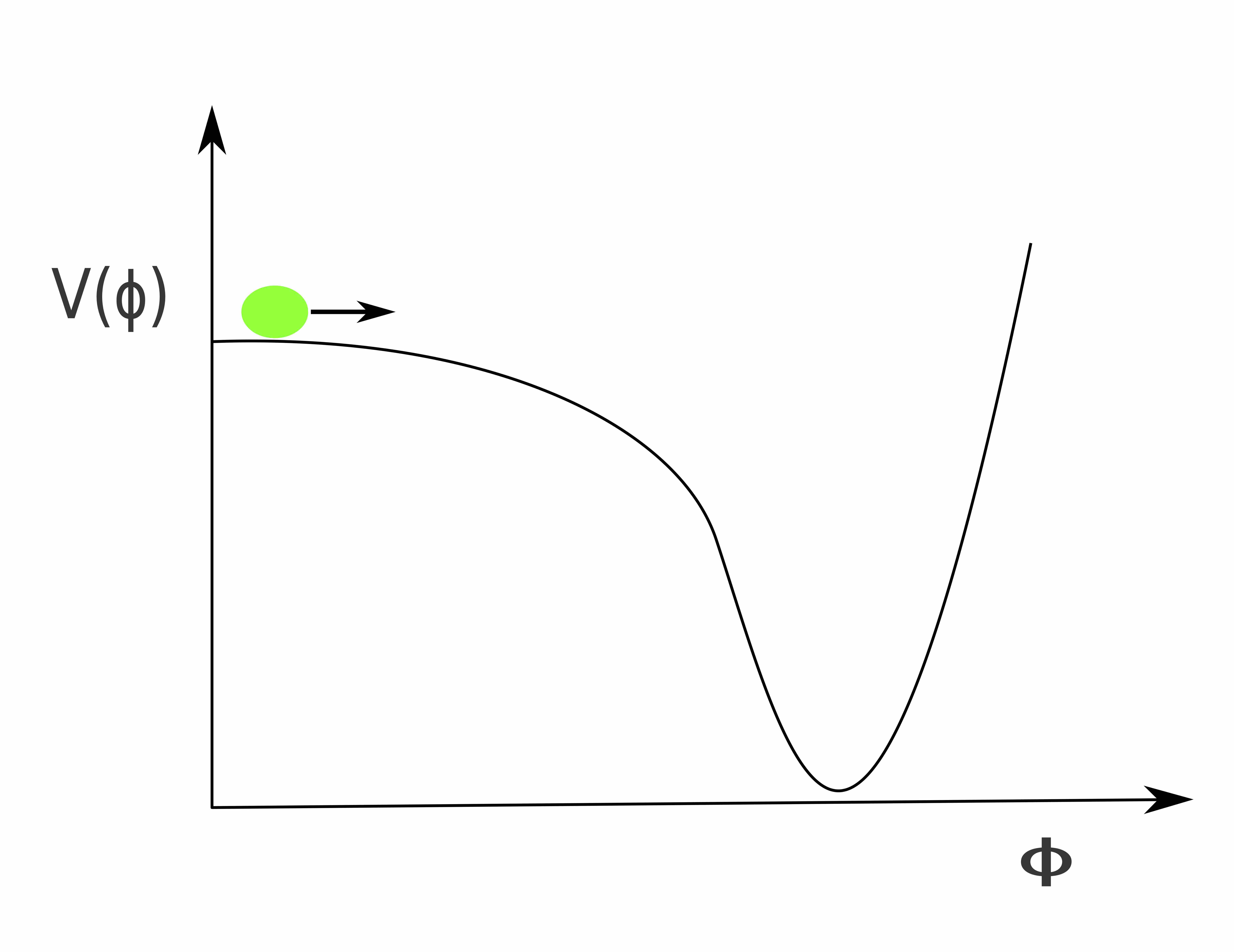}
	\caption{\footnotesize{New inflationary potential.}}
	\label{fig:new2}
 \end{center}	
\end{figure}

\textbf{Small field} models are typically described by potentials that arise naturally from spontaneous symmetry breaking. 
These types of models are also known as \textit{new inflation} \citep{Linde2, Parsons}. 
In this case, inflation takes place when the field is situated in a false vacuum state, very close to the top of the hill, and rolls 
down to a stable minimum, see Figure \ref{fig:new2}. These models are typically characterized by $V_{,\phi\phi} < 0$
and $\eta_{\rm v} < -\epsilon_{\rm v}$, usually $\epsilon_{\rm v}$  is closely zero (and hence the tensor amplitude). 

\noindent

Small field potentials can be written in \lp{a} generic form as

\begin{equation}
V\left(\phi\right) = \Lambda^4 \left[1 - \left(\phi / \mu\right)^p\right],
\end{equation}
where the exponent $p$ differs from model to model. $V(\phi)$ is usually considered as the lowest-order term in a Taylor expansion 
from a more general potential. In the simplest case of spontaneous symmetry breaking, with no special symmetries, 
the dominant term is the mass term, $p = 2$, hence the model gives

\bea
n_{\rm s}-1&\simeq& -\left( \frac{m_{\rm Pl}}{\mu}\right)^2, \nonumber \\
N  &=& \frac{4\pi\mu^2}{m_{\rm Pl}^2}\left[\ln\left(\frac{\phi_{end}}{\phi_i}\right)-\frac{\phi_{end}-\phi_i}{2\mu^2}\right],\nonumber\\
r &=& 8 (1 - n_{\rm s}) \exp\left[- 1 - N\left(1 - n_{\rm s}\right)\right].\\
\eea

On the other hand, $p > 2$ has a very different behavior. The scalar spectral index is
\begin{equation}
n_{\rm s}-1 = - {2 \over N} \left({p - 1 \over p - 2}\right),
\end{equation}
independent of $(m_{\rm Pl}/\mu)$. Besides, the tensor-to-scalar ratio for this model is given by 
\begin{equation}
 r = 8\left(\frac{\sqrt{8\pi}\mu}{m_{\rm Pl}}\right)^{2p/(p-2)}\left(\frac{p}{2N(p-2)}\right)^{2(p-1)/(p-2)}.
\end{equation}

\subsection{Hybrid models: $0 < \epsilon_{\rm v} < \eta_{\rm v}$}\label{hybridmodels}

The third class, called \textbf{hybrid models}, frequently includes those that incorporate supersymmetry into inflation 
\citep{Linde3, Copeland}.  In these models, the inflaton field $\phi$ evolves towards a minimum of its potential, however, the 
minimum has a vacuum energy  $V(\phi_{\rm min}) = \Lambda^4$ different from zero. 
In such cases, inflation continues forever unless an auxiliary field $\psi$ is added to interact with $\phi$ and ends inflation at 
some point $\phi = \phi_{\rm c}$.  Such models are well described by $V_{,\phi\phi} > 0$ and 
$0 < \epsilon_{\rm v} < \eta_{\rm v}$, where $V$ is the effective 1-field potential for the inflaton. 
\\

The generic potential for hybrid inflation, in a similar way to large field and small field models,
is considered as
\begin{equation}
V\left(\phi\right) = \Lambda^4 \left[1 + \left(\phi / \mu\right)^p\right],
\end{equation}
where again $p$ is an exponent that differs from model to model. For $\left({\phi / \mu}\right)\gg 1$, the behavior of the large-field 
models is recovered. Besides that, when $\left({\phi / \mu}\right)\ll 1$, the dynamics is similar to small-field models, but
now the field is evolving towards a dynamical fixed point rather than away from it.  
Because the presence of an auxiliary field the number of {\it e}-folds is

\beq
N(\phi)\simeq \left( {p+1 \over p+2} \right)\left[ {1 \over \eta(\phi_c) }- {1 \over \eta(\phi)}\right]. 
\eeq

\noindent
For $\phi \gg \phi_c$, $N(\phi)$ approaches the value
\beq
N_{max}\equiv \left( {p+1 \over p+2} \right) {1 \over \eta (\phi_c)}.
\eeq
In general 
 \beq
 N = \frac{8\pi\mu^p}{pm_{\rm Pl}^2}\left[\frac{\phi_{end}^{2-p}-\phi_i^{2-p}}{2-p}+\frac{\phi_{end}^2-\phi_i^2}{2\mu^p}\right], \ \ \ \text{for $p\neq 2$},\\
 \eeq
 \beq
 N = \frac{8\pi\mu^p}{pm_{\rm Pl}^2}\left[\ln\left(\frac{\phi_{end}}{\phi_i}\right)+\frac{\phi_{end}^2-\phi_i^2}{2\mu^p}\right], \ \ \ \text{for $p= 2$},
 \eeq
 and therefore, the spectral index is given by 
 $$
n_{\rm s}-1 \simeq 2 \left( \frac{p+1}{p+2}\right) \frac{1}{N_{max}-N}.
 $$
As we can note, the power spectrum is \textit{blue} ($n_{\rm s}>1$) and the model presents a running of the spectral index

\begin{equation}
\label{eq:dsirunning}
{dn_{\rm s} \over d\ln{k}} = -{1 \over 2} \left({p + 2 \over p + 1}\right) 
\left(n_{\rm s} - 1\right)^2.
\end{equation}
 This parameter  will be very useful for higher orders and more accurate constraints in  future observations. For instance, the particular 
 case $p = 2$ and $n_{\rm s} = 1.2$, the running obtained is $dn_{\rm s} / d\ln{k} = -0.05$ \citep{Kinney3}.

\subsection{Linear models: $\eta_{\rm v} = - \epsilon_{\rm v}$}

Linear models, $V\left(\phi\right) \propto \phi$, are located on the limits between
large field and small field models. They are represented by $V_{,\phi\phi} = 0$ and $\eta_{\rm v} =- \epsilon_{\rm v}$. 
The spectral index and tensor-to-scalar ratio are given by 

\beq
n_{\rm s}-1=-{6\over 1-4N},\qquad
r = {16 \over 1-4N} .
\eeq

\subsection{Logarithmic inflation}

There remain several single-field models which cannot fit into this classification, 
for instance, the logarithmic potentials \citep{Barrow2}

\beq
V\left(\phi\right) =V_0\left[1+(C g^2/8\pi^2)
\ln\left(\phi/\mu\right)\right].
\eeq
Typically they correspond to loop corrections in a supersymmetric theory, where $C$ denotes the degrees of freedom coupled
to the inflaton and $g$ is a coupling constant.  For this potential, the inflationary parameters are 
\bea
n_{\rm s}-1&\simeq&-\frac{1}{N}, \nonumber \\
r&\simeq& \sqrt{\frac{1}{N}{Cg^2\over 16\pi}}.
\eea
In this model, to end up inflation, an auxiliary field is needed, which is the main feature of hybrid models. However, when it is 
plotted on the $n_{\rm s}$---$r$ plane, it is located in the small-field region.
\\

\begin{figure}[t!]
\begin{center}
  \includegraphics[width=7cm, height=5.5cm]{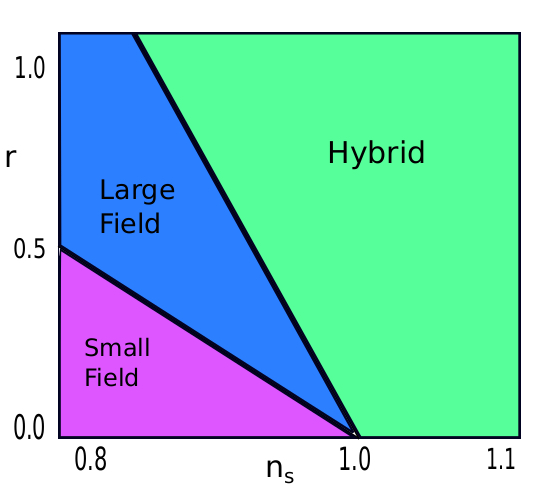}
	\caption{\footnotesize{Classification of the
potentials in terms of $n_{\rm s}$ and $r$ parameters.} }
\label{fig:parameters}
\end{center}
\end{figure}

\subsection{Hybrid Natural Inflation}

Hybrid Natural Inflation is particularly appealing because its origins lie in well motivated physics.
The inflaton potential relevant to the inflationary era has the general form
\begin{equation}
V(\phi)=\Delta^4(1+a\cos(\frac{\phi}{ f})),
\label{pot1}
\end{equation}
where $f$ is the symmetry breaking scale and $a$ allows for more general inflationary phenomena that can readily accommodate 
the Planck results, and even allow for a low-scale of inflation. Here the inflaton, $\phi$, is a pseudo-Goldstone boson 
associated with a spontaneously broken global symmetry and is thus protected from large radiative corrections to its mass.
Defining $c_{\phi}$ and $s_{\phi}$ by $\cos(\frac{\phi}{f})$ and $\sin(\frac{\phi}{f})$ respectively, we get
\begin{eqnarray}
\epsilon_{\rm v} &=&\frac{1}{16 \pi}\left(\frac{a}{f}\right)^2\frac{s_{\phi} ^{2}}{\left( 1+a\, c_{\phi} \right)^2} ,
\label{HNIeps}%
\\
\eta_{\rm v} &=&-\frac{1}{8\pi}\left( \frac{a}{f^2}\right)\, \frac{c_{\phi}}{1+a c_{\phi}} , 
\label{HNIzeta}
\end{eqnarray}%
and the inflationary parameters are computed and constrained by \citep{Vazquez4, Graham}.
\\

The classification of inflationary models mentioned previously may be interpreted as an 
arbitrary one, nevertheless, it is very useful because different types of models cover different 
regions of the $(n_{\rm s}, r)$ plane without overlapping, see Figure \ref{fig:parameters}.

\subsection{Hybrid waterfall inflation}

A two-field inflationary scenario  is an alternative case of the hybrid models. 
It occurs when the mass of the auxiliar field is smaller than the Hubble parameter, i.e.,  $V_{,\psi\psi}\lesssim H$. 
Once the inflaton acquires a critical value $\phi_c$, the auxiliary field starts evolving slowly, and 
a period of inflation is produced during its dynamics, usually called the \textit{waterfall scenario}. An interesting result is the 
possibility to obtain a \textit{red} power spectrum ($n_{\rm s}<1$), according to the amount of inflation produced during the 
waterfall period.   
As an example, let us consider two scalar fields with a potential $V_t$ like chaotic-hybrid:
\begin{equation}
V_t=\frac{\lambda}{4}\left[\left(\frac{M^2}{\lambda}-\psi^2\right)^2+\frac{1}{2}m^2\phi^2+\frac{1}{2}g^2\phi^2\psi^2\right],
\end{equation}
with $M, m, \lambda$ constant values.
In the typical hybrid models, it is expected that the \textit{waterfall} field $\psi$ remains at $\psi=0$ while the inflaton field $\phi$ 
evolves generating inflation. Then, when $\phi=\phi_c$, the minimum $\psi=0$ becomes unstable, and the waterfall 
field rolls down to its true minimum, finishing up immediately with the inflationary era. 
However, if $M^2\lesssim H^2$, we obtain the waterfall period. Taking the  limit $g^2\psi^2/H^2\ll m^2/H^2$ (i.e., the 
back-reaction of the waterfall field on the inflaton is small during inflation) and $\psi^2/H^2\ll M^2/\lambda H^2$ we obtain 
finally that \citep{Abol}
\begin{equation}
n_{\rm s}-1\simeq \left[\frac{4M^2}{3H^2}\left(\frac{M^2}{9H^2}-rn_k\right)\right]_{k=aH},
\end{equation}
where $n_k=N_k-N_c$ is a measurement of the difference between the $e$-folds $N_k$  
 when a given scale $k$ has left the horizon and the $e$-folds $N_c$ when the waterfall transition starts. 
 Then, for modes that left the horizon before the phase transition, we have $n_k<0$ and $n_{\rm s} >1$, 
 whereas, for modes that have left the horizon after a phase transition, we have that $n_k>0$ and $n_{\rm s}$ can take any value.

\section{Observational results}

 \begin{figure}[t]
 \begin{center}
  \includegraphics[trim = 1mm 1mm 1mm 1mm, clip, width=5.7cm, height=3.5cm]{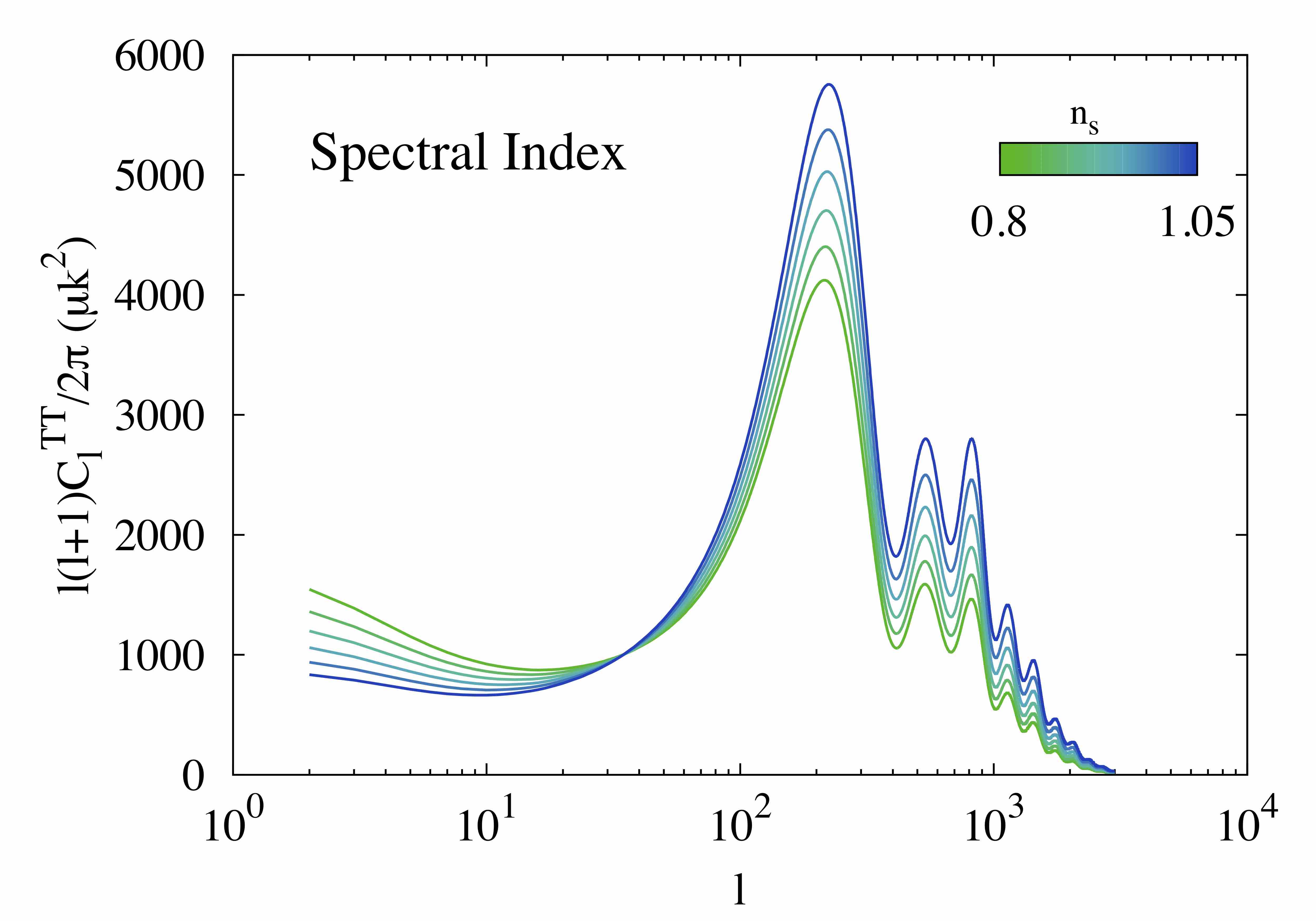}
  \includegraphics[trim = 1mm 1mm 1mm 1mm, clip, width=5.7cm, height=3.5cm]{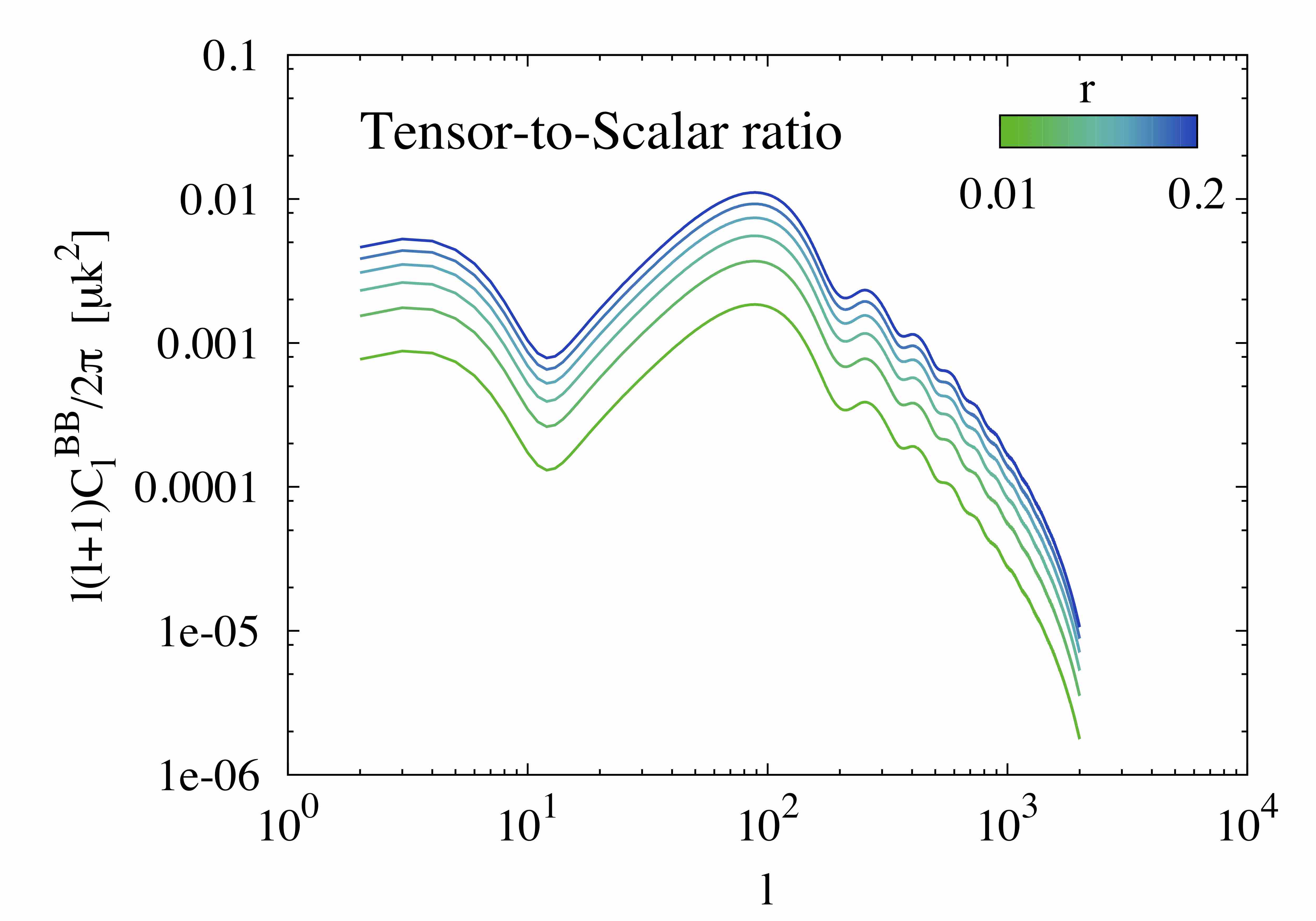}
	\caption{\footnotesize{Variations
	of the CMB scalar spectrum for different values of the spectral index $n_{\rm s}$ (left), 
	and variations of the CMB tensor spectrum with respect to the tensor-to-scalar ratio $r$ (right).}}
	\label{fig:CMB_spectra}
 \end{center}	
\end{figure}

How can observations constrain $n_{\rm s}$ and $r$ in inflationary models?
During several years many projects, at different scales, have been carried out to look for observational data to constrain 
cosmological models. That is, different models may imprint different behaviors over the CMB spectra, 
see Figure \ref{fig:CMB_spectra}. Amongst many projects, they are:  Cosmic Background Explorer (COBE), Wilkinson Microwave 
Anisotropy Probe (WMAP), Cosmic Background Imager observations (CBI), Ballon Observations of Millimetric Extra-galactic 
Radiation and Geophysics (BOOMERang), the Luminous Red Galaxy (LRG) subset DR7 of the Sloan Digital Sky Survey (SDSS), 
Baryon Acoustic Oscillations (BAO), Supernovae (SNe) data, Hubble Space Telescope (HST) and recently the South Pole 
Telescope (SPT), the Atacama Cosmology Telescope (ACT) and the Planck Satellite. Below, we show some of the constraints for 
different types of inflationary potentials by using historical and current observational data.
We stress that the results are shown on the phase space $n_{\rm s}-r$, and therefore our interest is mainly focussed 
on the case with no running $dn_{\rm s}/d\ln k =0$ and single fields.
 \\

\begin{figure}[t!]
\begin{center}
  \includegraphics[trim = 1mm 1mm 3mm 10mm, clip, width=4cm, height=4.5cm]{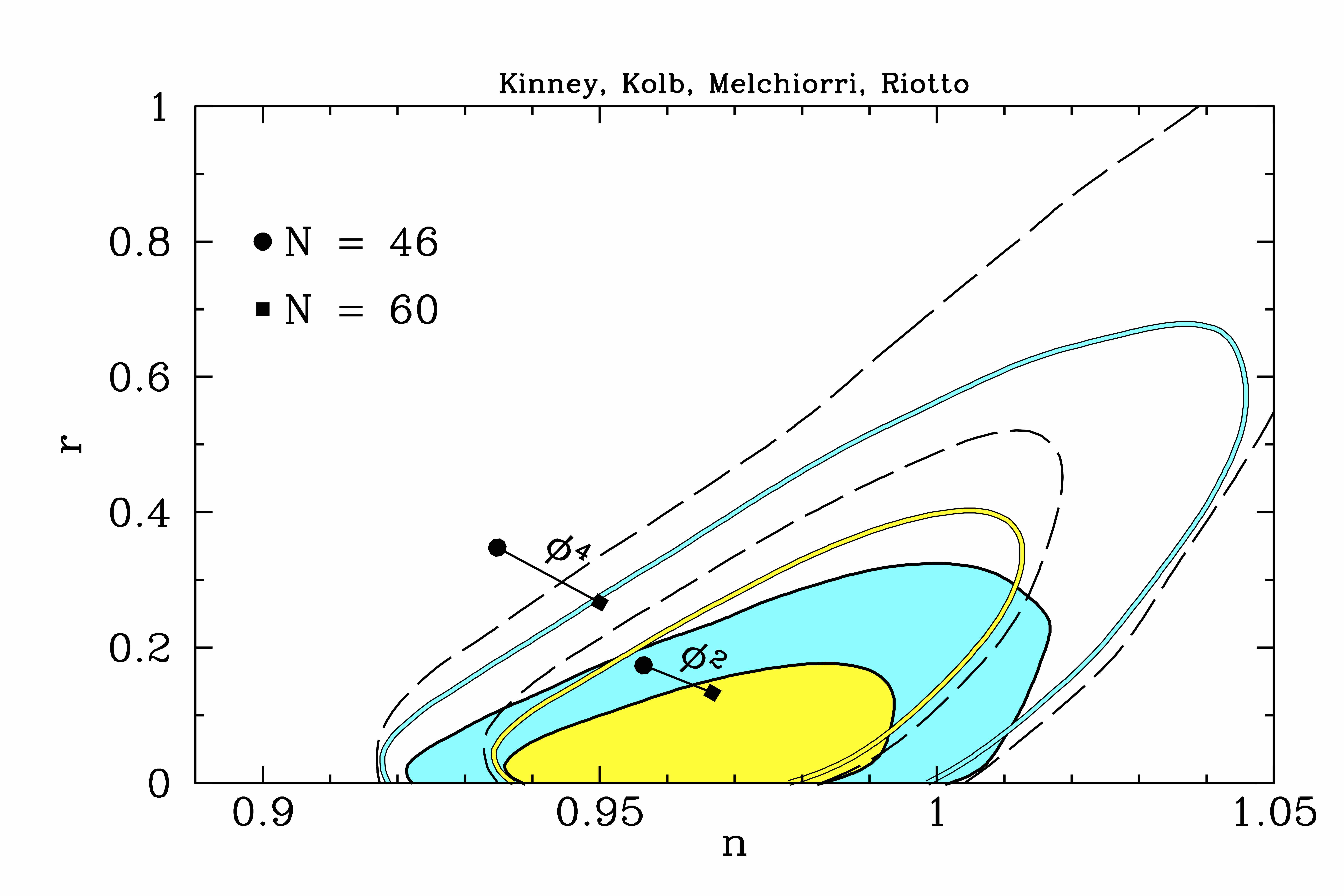} 
\caption{ 
\footnotesize{WMAP3 only (open contours) and 
WMAP3+SDSS (filled contours) 2D posterior distributions on the phase space $n_{\rm s}$-$r$,
for the potentials $\phi^2$ and $\phi^4$ by considering  $e$-folds of $N\sim$ 46 and 60.
Colored regions correspond to 68\% and 95\% CL~\citep{Kinney}.
}}
\label{fig:Kinney}
\end{center}
\end{figure}

Figure \ref{fig:Kinney} displays 2D marginalized posterior distributions for $n_{\rm s}$ and $r$ based on two data sets: WMAP3 by itself, 
and WMAP3 plus information from the LRG subset from SDSS \citep{Kinney}. Considering WMAP3 observations alone (open contours)
the parameters are constrained such that $0.94 < n_{\rm s} < 1.04$ and $r<0.60$ (95\% CL). Those models that present  
$n_{\rm s}<0.9$ are  therefore ruled out at high confidence level. The same is applied for models with $n_{\rm s} > 1.05$. 
WMAP data by itself  cannot lead to strong constraints, because of the existence of parameter degeneracies, like the well known 
geometrical degeneracy involving $\Omega_m$,  $\Omega_{\Lambda}$ and $\Omega_k$. However, when it is combined with 
different types of datasets, together, they increase the constraining power and might remove degeneracies. 
Once the SDSS data is included, the limit of the gravitational wave amplitude and the spectral index constraints are  reduced, that is, 
for WMAP3+SDSS (filled contours) the constraints on $n_{\rm s}$ and $r$ are $0.93<n_{\rm s}<1.01$ and $r<0.31$. 
Moreover, Figure \ref{fig:Kinney} shows that the Harrison-Zel'dovich model: $n_{\rm s}=1, r=0, ~dn_{\rm s}/d \ln k=0$,
is still in  good agreement with this type of data.
Similarly, for inflation driven by a massless self-interacting scalar field $V(\phi) = \lambda\phi^4$ (see equation \eqref{nsr}), 
the contours indicate that this potential with 60 \textit{e}-folds is still consistent with WMAP3 data at 95\% CL,
nevertheless ruled out by the combined datasets WMAP3+SDSS.
The potential $V(\phi) = m^2\phi^2/2$ is consistent with both data sets, with a preference to 60 $e$-folds.
\\

\begin{figure}[t!]
\begin{center}
 \includegraphics[trim = 1mm 10mm -2mm -10mm, clip, width=3.cm, height=4.cm]{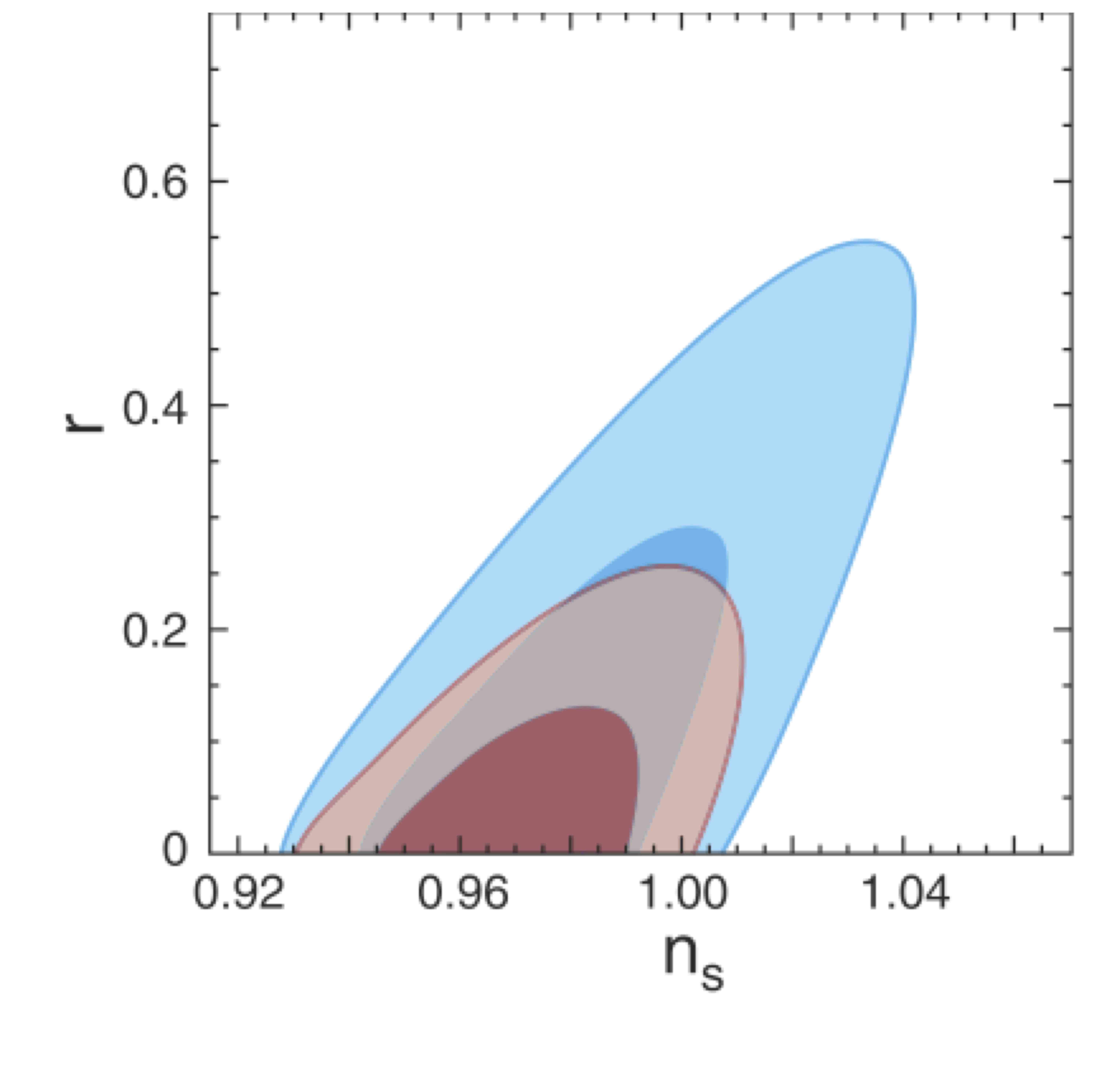}
  \includegraphics[trim = 0mm 0mm 2mm 0mm, clip, width=8.6cm, height=4.2cm]{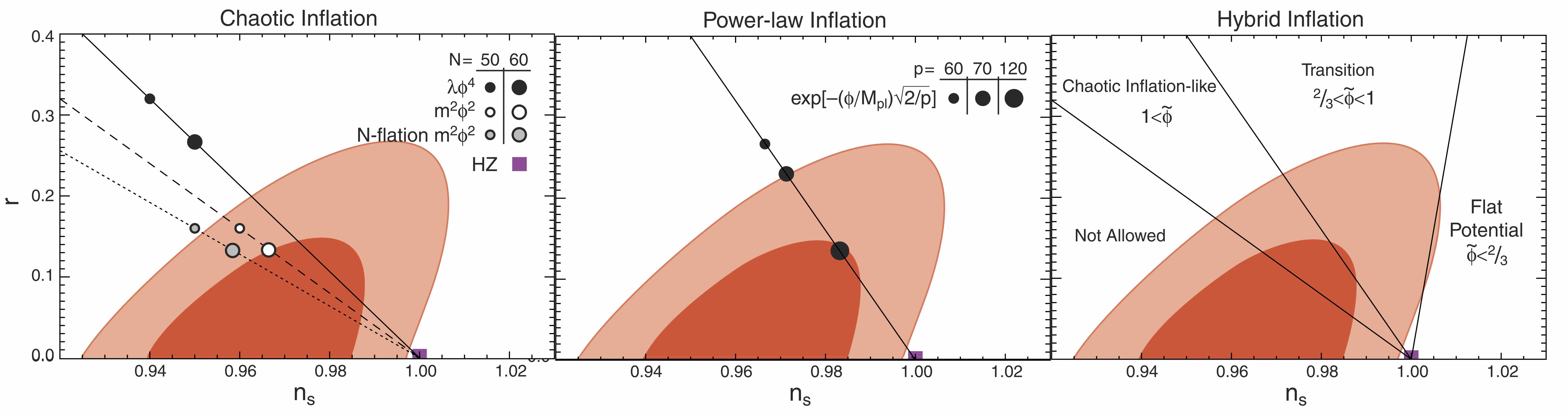} 
\caption{\footnotesize{Constraints on $n_{\rm s}$ and $r$.
Left panel: WMAP5 results are colored blue and WMAP5+BAO+SN red.
Right panel: Constraints on large and hybrid models from the combined datasets WMAP5+BAO+SN.
Colored regions correspond to 68\% and 95\% CL
 \citep{Komatsu}.}}\label{fig:Komatsu}
\end{center}
\end{figure}

On the other hand, left panel of figure \ref{fig:Komatsu} shows limits imposed by WMAP5 data alone,  $r < 0.43$ (95\% CL) 
while $0.964< n_{\rm s}<1.008$. When BAO and SN data are added, the limits improved significantly to 
$r < 0.22$ (95\% CL) and {$0.953< n_{\rm s}<0.983$ \citep{Komatsu}}. Right panel of figure \ref{fig:Komatsu} displays a summary 
for different potential constraints by WMAP5+BAO+SN.
The model $V(\phi)=\lambda \phi^4$, unlike WMAP3 constraints, is found to be located far away from the 95\% CL, and 
therefore it is excluded by more than 2$\sigma$. For inflation produced by a massive  scalar field $V(\phi)=(1/2)m^2\phi^2$, the 
model with $N=50$ is situated outside the 68\% CL, whereas with $N=60$ is at the boundary of the 68\% CL. 
Therefore, this model is consistent with data within the 95\% CL. 
The points represented by $N$-inflation describe a model with many massive axion fields \citep{Liddle3}. 
For an exponential potential $\left(V(\phi)=\exp\left[-(\phi/m_{pl})\sqrt{2/p}\right]\right)$, it is observed that models 
with $p<60$ are mainly excluded. Models with $60<p<70$ are roughly in the boundary of the 95\% region, and $p>70$ are 
in agreement within the 95\% CL.  Some models with $p\sim 120$ essentially layout in the limit of the 68\% CL.
\\

The hybrid potentials, as already noted, can have different behaviors
depending on the $(\phi/ \mu)$ value. The parameter space can be split up into three 
different regions based on $(\phi / \mu)$. For $\phi / \mu \ll 1$ the dynamics is similar to small
fields and the dominant term lays in the region called Flat Potential Regime. 
For $\phi / \mu \gg 1$, the results are similar to large field models, and this region is called
Chaotic Inflation-like Regime. The boundary, $\phi / \mu \sim 1$ is named 
Transition regime. The different $(\phi / \mu)$ values corresponding to their regions
are shown in the right panel of Figure \ref{fig:Komatsu}.
Finally, the combined datasets WMAP5+BAO+SN ruled out the Harrison-Zel'dovich model
by more than 95\% CL.
\\

\begin{figure}[h!]
\begin{center}
 \includegraphics[trim = 10mm 50mm 20mm 40mm, clip, width=7cm, height=5cm]{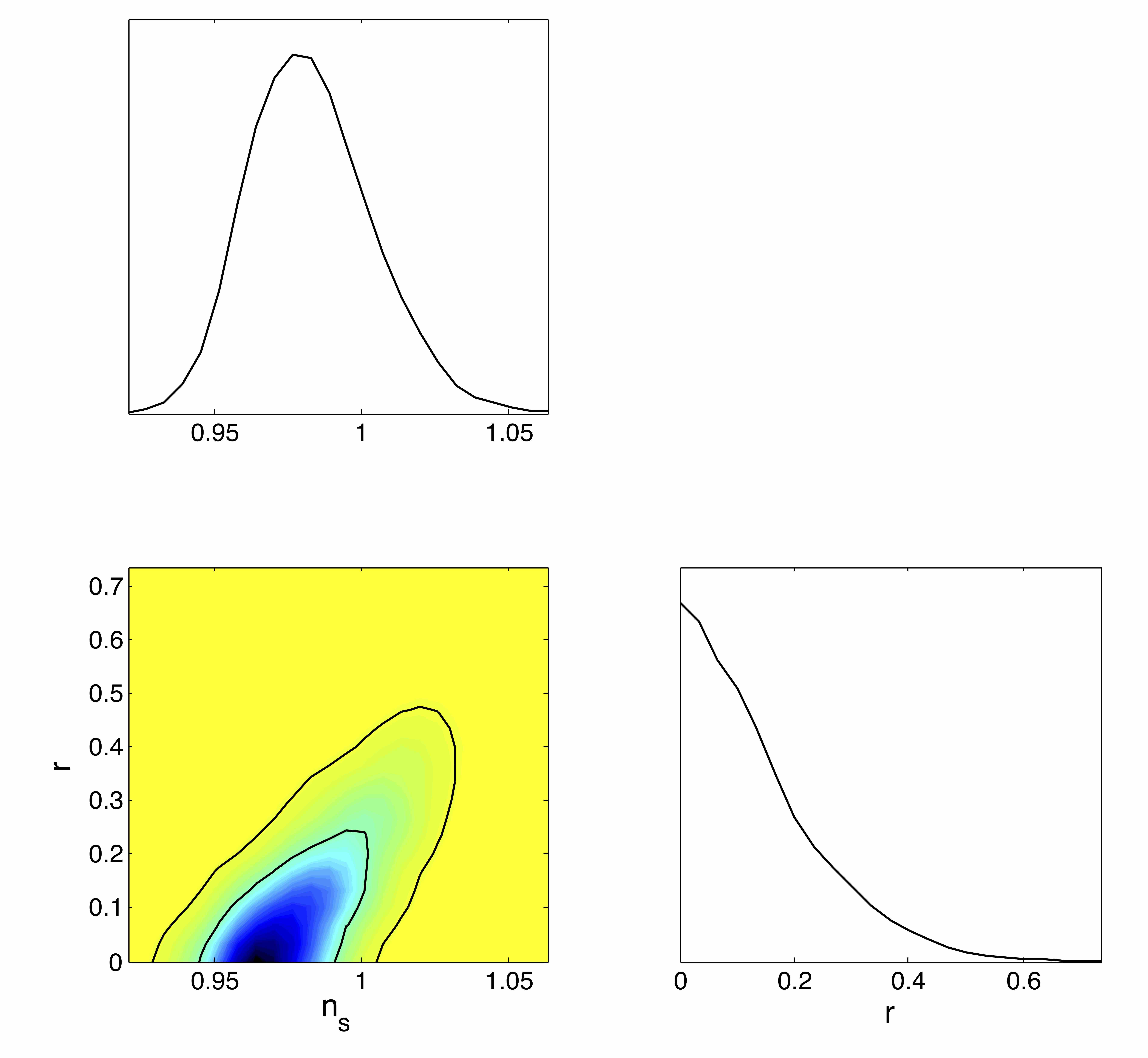}
\caption{\footnotesize{1D and 2D Marginalized probability constraints on $n_{\rm s}$ and $r$ using only WMAP7 data. 
2D constraints are plotted with $1\sigma$ and $2\sigma$ confidence contours.
}}\label{fig:infla}
\end{center}
\end{figure}

Following the same line for inflationary models, we use the {\sc cosmoMC} package \citep{Lewis}
which allows to perform the parameter estimation and  provide constraints for the $n_{\rm s}$ and $r$ parameters, given 
a dataset [we refer to \citet{Padilla} where the authors provided an introduction on Bayesian parameter inference and 
its applications to cosmology]. We assume a flat $\Lambda$CDM model 
specified by the following parameters: the physical baryon $\Omega_{\rm b} h^2$ and cold dark matter density
 $\Omega_{\rm DM } h^2$ relative to the critical density, $\theta$ is $100 \times$ the ratio of the sound horizon 
 to angular diameter distance at last scattering surface and $\tau$ denotes the optical depth at reionization.
To illustrate our point, we initially consider WMAP seven-year data. We observe from Figure \ref{fig:infla} that a 
model to be considered as a favorable candidate it has to predict a spectral index about  $n_{\rm s}=0.982^{+0.020}_{-0.019}$ 
 and a tensor-to-scalar ratio $r<0.37$ (95\% CL). When WMAP-7 is combined with different datasets, the constraints are 
 tightened, as it is shown by \citet{Larson}.

\begin{figure}[h!]
\begin{center}
 \includegraphics[trim = 1mm 1mm 1mm 1mm, clip, width=6cm, height=5cm]{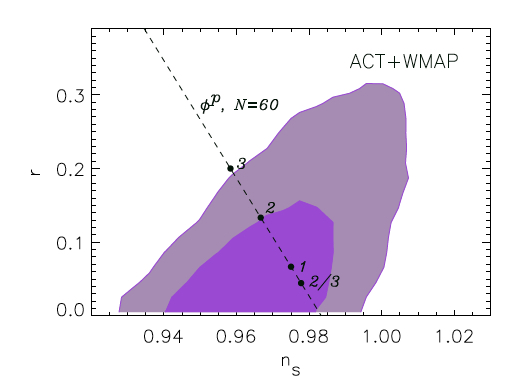}
  \includegraphics[trim = 1mm 1mm 1mm 1mm, clip, width=6cm, height=5cm]{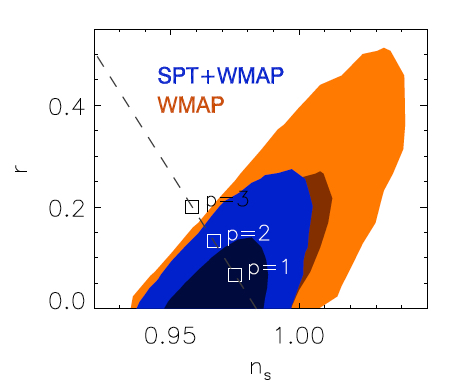}
\caption{\footnotesize{Marginalized 2D probability distribution (68\% and 95\% CL) for the 
tensor-to-scalar ratio $r$, and the scalar spectral index $n_{\rm s}$ for ACT+WMAP (left panel)
and SPT+WMAP (right panel)
 \citep{ACT,SPT}.}}
 \label{fig:SPT}
 \end{center}
\end{figure}

Two recent experiments have placed new constraints on the cosmological parameters: the Atacama
Cosmology Telescope (ACT) \citet{ACT} and the South Pole Telescope (SPT) \citet{SPT}.
Figure \ref{fig:SPT} shows the predicted values for a chaotic inflationary model with inflaton
potential $V(\phi)\propto\phi^p$ with 60 $e$-folds. We observe that models with $p\ge3$
are disfavored at more than 95\% CL.
\\

\begin{figure}[h!]
 \begin{center}
  \includegraphics[trim = 1mm 1mm 1mm 1mm, clip, width=9cm, height=5cm]{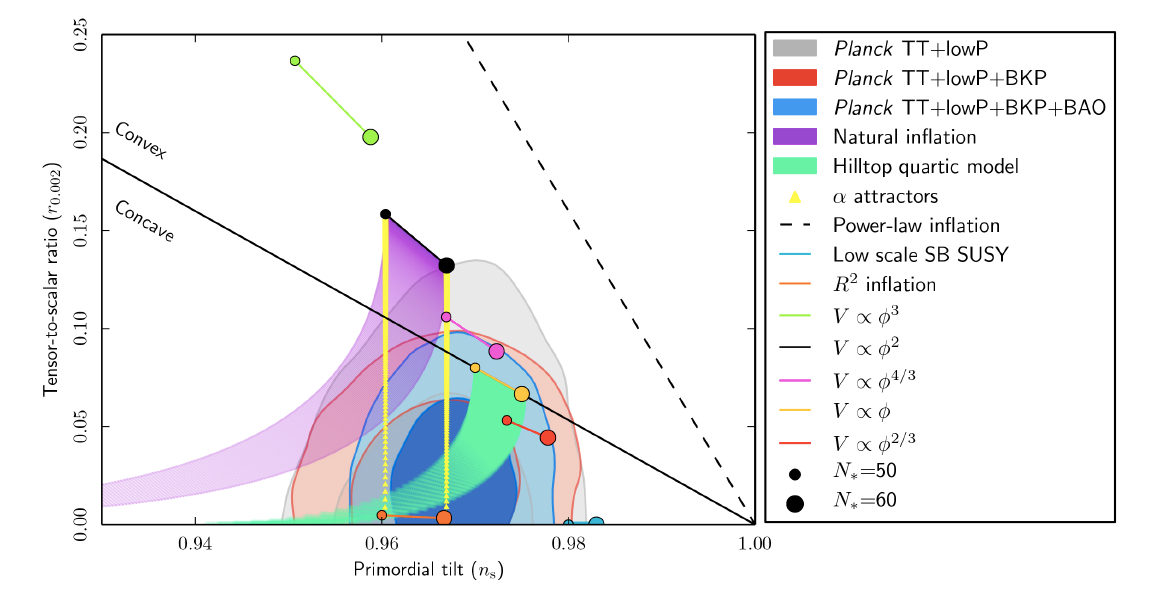}
\caption{\footnotesize{2D marginalized probability constraints on $n_{\rm s}$ and $r$ for
the most recent results of \citep{PlanckC}. 2D constraints are plotted with $1\sigma$ and $2\sigma$
confidence contours. The figure is taken from \citet{PlanckC}.
}}\label{fig:infla_2}
\end{center}
\end{figure}

\begin{figure}[h!]
 \begin{center}
  \includegraphics[trim = 1mm 1mm 1mm 1mm, clip, width=7cm, height=5cm]{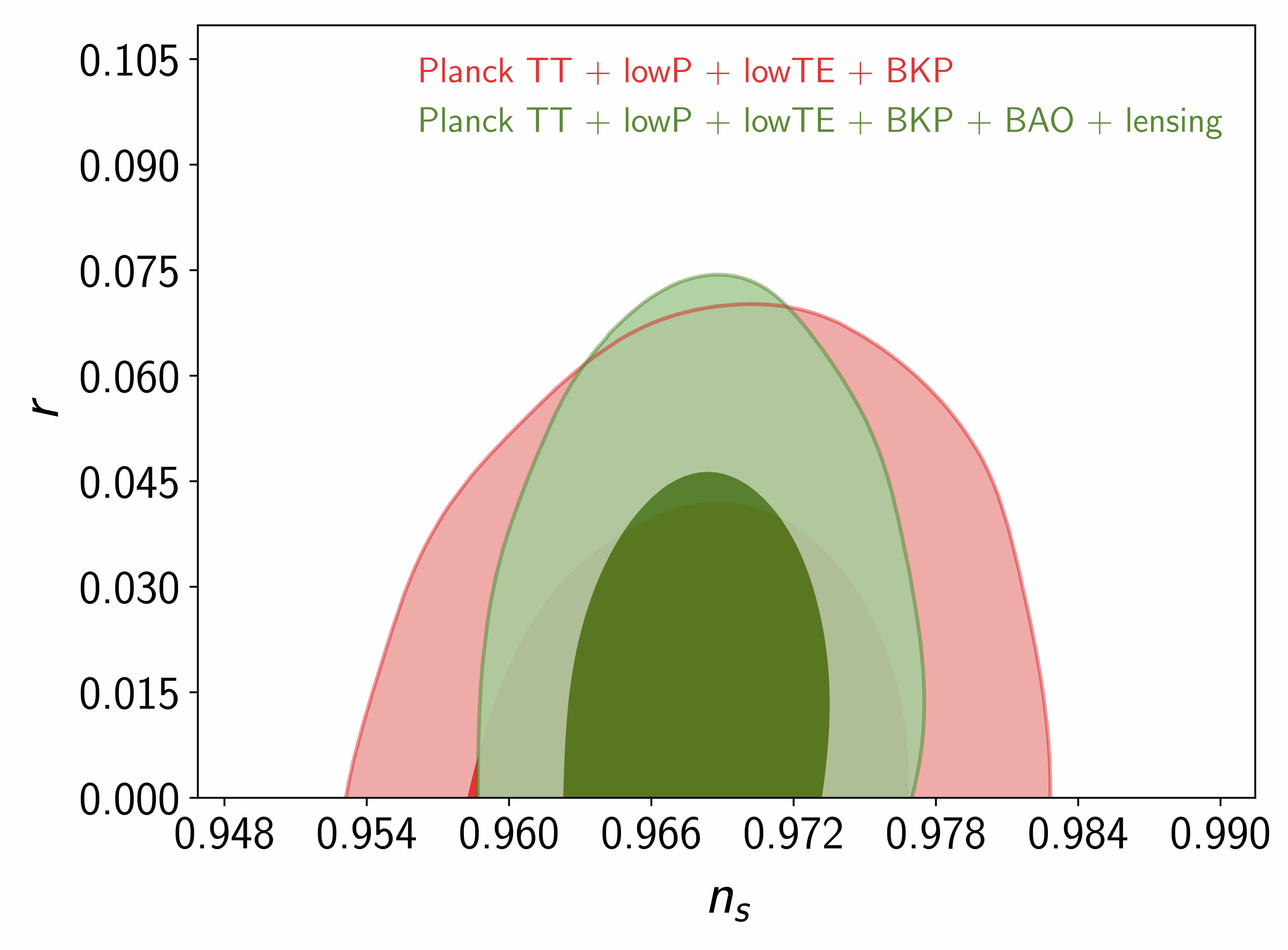}
    \includegraphics[trim = 1mm 1mm 1mm 1mm, clip, width=7cm, height=5cm]{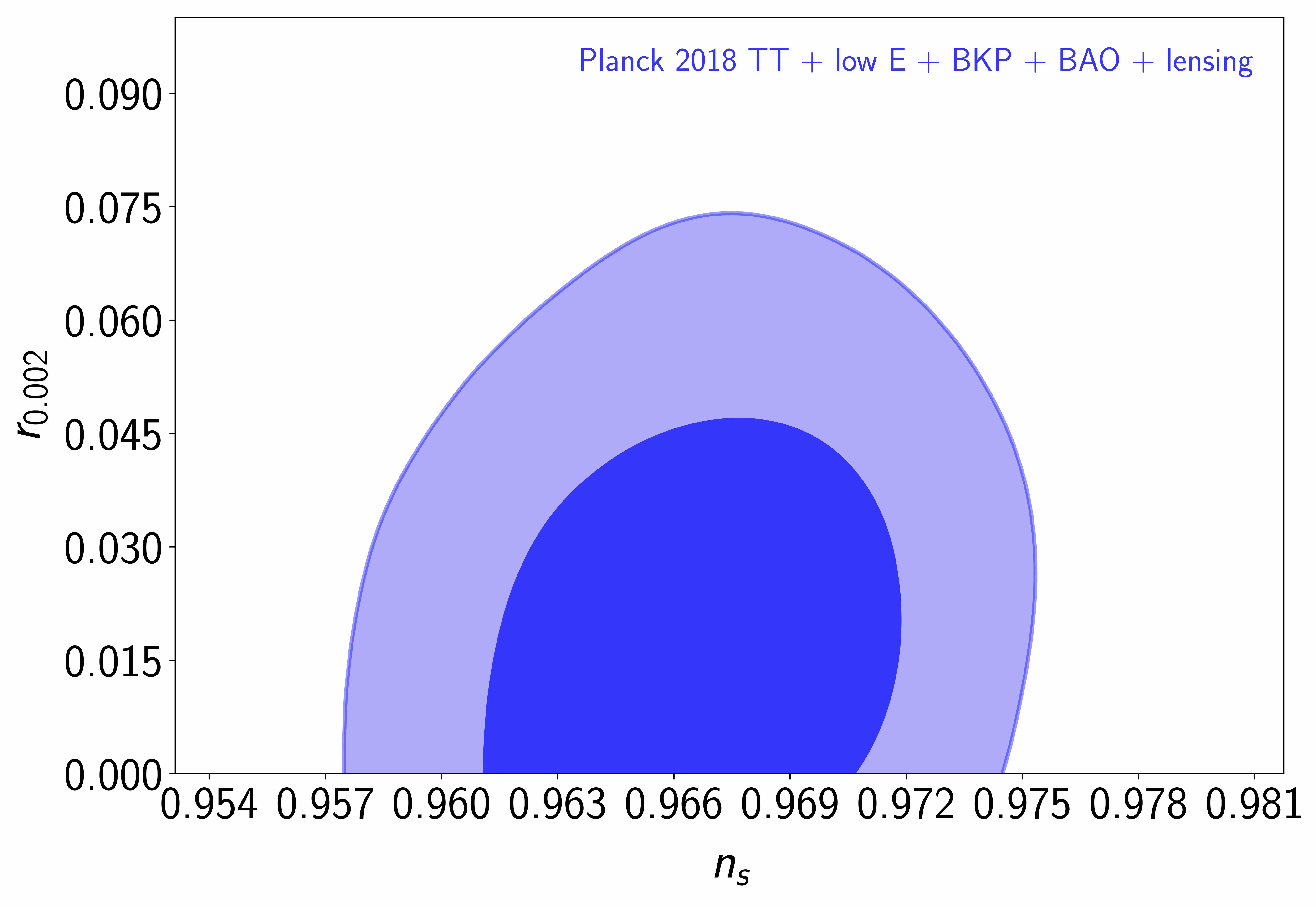}
\caption{\footnotesize{2D marginalized probability constraints on $n_{\rm s}$ and $r$ for
the Planck 2015 dataset (left) and Planck 2018 (right). 2D constraints are plotted with $1\sigma$ and $2\sigma$
confidence contours. The figure was done by using the CosmoMC package.
}}\label{fig:infla_3}
\end{center}
\end{figure}

Figure \ref{fig:infla_2}  shows recent constraints given by \citet{PlanckC} in the $n_{\rm s}$ and $r$ plane. 
Gray regions correspond to the Planck 2013 results, red regions added the contribution of the temperature 
power spectrum (TT) and the Planck polarization data in the low-$l$ likelihood (lowP) while blue regions added 
the temperature-polarization cross spectrum (TE), and the polarization power spectrum (EE).
Notice that the model that fits the best to the data corresponds to $R^2$ inflation \citep{Starobinsky},
and models $V(\phi)\propto\phi^p$ with $p\geq 2$ are discarded by data.
The addition of BAO data and lensing is shown in the left panel of Figure \ref{fig:infla_3}. 
 Finally, to incorporate the most updated version of the data, on the right panel of Figure \ref{fig:infla_3}, 
 we include into the CosmoMC code  the full-mission \textit{Planck 2018 }
 (TT,TE,EE+lowE+lensing)  \citep{Planck18}, the  Keck Array, and BICEP2 Collaborations 2016 \citep{BKP16} 
 and the BAO data \citep{BAO2}  in order to tighten the parameter space constraints.

\section{Conclusions}

\begin{table}[h!]\centering
\begin{tabular}{|c|c|c|}
\toprule
Parameter & Limits & Data set\\
\hline
$n_{\rm s}$& $ 0.9661 \pm {0.0037}$ & Planck 2018 TT + low E + BKP  \\ 
$r$ & $< 0.065$ & + BAO + lensing\\
\hline
$n_{\rm s}$& $ 0.9683 \pm {0.0059}$ & Planck TT + lowP + lowTE + BKP \\ 
$r$ & $< 0.0660$ & + BAO + lensing\\
\hline
$n_{\rm s}$& $ 0.9666 \pm {0.0062}$ & Planck TT+lowP\\ 
$r$ & $< 0.103$ & \\
\hline
$n_{\rm s}$& $ 0.9711 \pm {0.0099}$ & SPT+WMAP7+BAO+$H_0$\\ 
$r$ & $< 0.17$ & \\
\hline
$n_{\rm s}$& $ 0.970 \pm {0.012}$ & ACT+WMAP7+BAO+$H_0$\\ 
$r$ & $< 0.19$ & \\
\hline
$n_{\rm s}$& $ 0.973 \pm 0.014$ & WMAP7 + BAO +$H_0$\\ 
$r$ & $< 0.24$ & \\
\hline
$n_{\rm s}$& $ 0.982 \pm ^{+0.020}_{-0.019}$ & WMAP7 ONLY\\ 
$r$ & $< 0.36$ & \\
\hline
$n_{\rm s}$& $ 0.968 \pm 0.015$ & WMAP5+BAO+SN\\ 
$r$ & $< 0.22$ & \\
\hline
$n_{\rm s}$ & $0.986\pm  0.022$ & WMAP5 ONLY  \\
$r$ & $  < 0.43 $ &  \\
\hline
$n_{\rm s}$& $0.97\pm 0.04 $ & WMAP3 + SDSS\\ 
$ r$& $<0.31$ & \\
 \hline
$n_{\rm s}$& $0.99 \pm 0.05 $ & WMAP3 ONLY \\
 $r$& $< 0.60 $ &  \\
\bottomrule
\end{tabular}
\caption{Summary of the \lowercase{$n_{\rm s}$, $r$} constraints from different measurements 
\citep{Peiris et al.2003; Kinney et al.2006; 
Komatsu et al.2009; Komatsu et al.2011; Dunkley et al. 2010; Keisler et al. 2011, Planck}}
\label{tab:resul}
\end{table}

Considering the analysis presented here, it is complicated to prove that a given model is correct, since these models could 
be just particular cases of more general scenarios with several parameters involved. However, it is possible to eliminate models 
or at least give some constraints on their behavior, leading to a narrower range of study.
Although we have presented some simple examples of potentials, the classification in small-field, large-field, and hybrid models 
is enough to cover the entire region of the $n_{\rm s}$--$r$ plane, as illustrated in Figure \ref{fig:parameters}.  
Different versions of the three types of models predict qualitatively different scalar and tensor spectra, so it should be particularly 
easy to work on them apart.
\\

We have seen that the favored models are those with small $r$ (assuming $dn_{\rm s}/d\ln{k}\sim 0$) and slightly \textit{red} 
spectrum, hence models with \textit{blue} power spectrum  $n_{\rm s} > 1.0$ are inconsistent with the recent data. These simple 
but important constraints allow us to rule out the simplest models corresponding to hybrid inflation of the form 
$V(\phi) = \Lambda^4(1 + (\mu / \phi)^{p})$. There remain models with red spectra in  the hybrid classification: inverted 
models and models with logarithmic potentials. 
\\

Table \ref{tab:resul} summarizes the constraints on the $n_{\rm s}$ and $r$ parameters and their improvements through the years.
The scale-invariant power spectrum $n_{\rm s} = 1$ is consistent within 95\% CL with WMAP3 data,  and therefore, not ruled out; 
however, with WMAP5 data the HZ spectrum lays outside the 95\% CL region, which indicates exclusion considering the lowest order 
on the $n_{\rm s}, r$ parameters. When WMAP7 data is considered, the scale-invariant spectrum is totally excluded by more 
than $3 \sigma$; however, the inclusion of extra parameters in a particular model may weaken the constraints on the spectral index.
When chaotic models $V(\phi)\propto\phi^p$ are analyzed with current data, it is found that quartic models 
($p=4$) are ruled out, whilst models with $p\ge3$ are disfavored at $>$ 95\% CL. Moreover, the quadratic potential 
$V(\phi)= 1/2 m^2 \phi^2$ is in agreement with all data sets presented here and therefore remains as a good candidate.
Future surveys will  provide a more accurate description of the universe, and therefore, 
narrow down the number of candidates, which might better explain the inflationary period.

\section{Acknowledgments }

LEP was supported by CONACyT M\'exico.
J.A.V. acknowledges the support provided by FOSEC SEP-CONACYT Investigaci\'on B\'asica A1-S-21925, and UNAM-DGAPA-PAPIIT IA102219

\bibliography{Inflation}


\end{document}